\documentclass[12pt]{article}
\usepackage{amsfonts}
\usepackage{amsmath,amssymb}
\usepackage{mathrsfs}
\usepackage{amsxtra}
\usepackage{amstext}
\usepackage{amscd}
\usepackage{latexsym}
\usepackage{bbold}
\newtheorem{thm}{Theorem}[section]
\newtheorem{prop}[thm]{Proposition}
\newtheorem{lem}[thm]{Lemma}
\newtheorem{cor}[thm]{Corollary}
\newtheorem{remark}[thm]{Remark}

\numberwithin{equation}{section}

\def\Tr{{\rm Tr}}

\def\Bbb{\mathbb}
\def\R{\Bbb R}
\def\C{\Bbb C}

\def\N{\Bbb N}

\usepackage{color}

\newcommand{\red}{\textcolor{red}}

\newcommand{\thedate}

\begin{document}
\hfill \textbf{\red{DOSfv}}

\[
\mbox{\bf\large Dynamics of an Open System for Repeated}
\]
\[
\mbox{\bf\large  Harmonic Perturbation}
\]

\vskip1cm

\begin{center}

\setcounter{footnote}{0}
\renewcommand{\thefootnote}{\arabic{footnote}}

\textbf{Hiroshi Tamura} \footnote{tamurah@staff.kanazawa-u.ac.jp}\\
      {Institute of Science and Engineering}      \\
              and \\
        Graduate School of the Natural Science and Technology\\
           Kanazawa University,\\
          Kanazawa 920-1192, Japan

\bigskip

\textbf{Valentin A.Zagrebnov }\footnote{Valentin.Zagrebnov@univ-amu.fr}\\
Institut de Math\'{e}matiques de Marseille - UMR 7373 \\
CMI-AMU, Technop\^{o}le Ch\^{a}teau-Gombert\\
39, rue F. Joliot Curie, 13453 Marseille Cedex 13, France\\
and \\
D\'{e}partement de Math\'{e}matiques \\
Universit\'{e} d'Aix-Marseille - Luminy, Case 901\\
163 av.de Luminy, 13288 Marseille Cedex 09, France

\vspace{1.5cm}

ABSTRACT

\end{center}

We use the Kossakowski-Lindblad-Davies formalism to consider an open system defined as the Markovian extension of one-mode
quantum oscillator $\mathcal{S}$, which is perturbed by a piecewise stationary harmonic interaction with a chain of oscillators
$\mathcal{C}$. The long-time asymptotic behaviour of various subsystems of
$\mathcal{S} + \mathcal{C}$ are obtained in the framework of the dual $W^*$-dynamical system approach.

\newpage
\tableofcontents
\section{Introduction} \label{Intr}
A quantum Hamiltonian system with time-dependent repeated harmonic interaction was proposed and investigated in \cite{TZ}.
The corresponding open system can be defined through the Kossakowski-Lindblad-Davies \textit{dissipative} extension of
the Hamiltonian dynamics. In our previous paper \cite{TZ1} the existence and uniqueness of the evolution map for density
matrices of the open system are established and its dual $W^*$-dynamics on the CCR $C^*$-algebra was described explicitly.

The aim of this paper is to apply the formalism developed in \cite{TZ1} to
analysis of dynamics of subsystems,
including their long-time asymptotic behaviour and correlations.

Let  $a$ and $a^*$ be the annihilation and the creation operators defined in the
Fock space $\mathscr{F}$ generated by a cyclic vector $\Omega\,(vacuum)$.
That is, the Hilbert space $\mathscr{F}$ is the completion of the algebraic span
$\mathscr{F}_{\mbox{\tiny fin}}$ of vectors $\{(a^{*})^m \Omega\}_{m\geqslant 0}$
and $a, a^*$ satisfy the Canonical Commutation Relations (CCR)
\begin{equation}\label{CCR0}
      [a, a^*] = \mathbb{1}, \quad  [a, a] = 0, \quad [a^*, a^*] = 0
      \quad \mbox{on} \quad \mathscr{F}_{\mbox{\tiny fin}}.
\end{equation}
We denote by $\{\mathscr{H}_k\}_{k=0}^{N}$ the copies of $\mathscr{F}$ for an arbitrary but finite $N \in \mathbb{N}$ and by
$\mathscr{H}^{(N)} $ the Hilbert space tensor product of these copies:
\begin{equation}\label{H-space}
          \mathscr{H}^{(N)}  := \bigotimes_{k=0}^{N} \mathscr{H}_k
                = \mathscr{F}^{\otimes (N+1)}  \ .
\end{equation}
In this space we define for $k =0, 1, 2, \ldots, N$ the operators
\begin{eqnarray}\label{k-bosons}
b_k :=
\mathbb{1} \otimes  \ldots \otimes \mathbb{1} \otimes a \otimes \mathbb{1} \otimes  \ldots \otimes \mathbb{1} \ ,  \ \ \
b_{k}^* :=
\mathbb{1} \otimes  \ldots \otimes \mathbb{1} \otimes a^* \otimes \mathbb{1} \otimes  \ldots \otimes \mathbb{1} \ ,
\end{eqnarray}
where operator $a$ (respectively $a^*$) is the $(k+1)$th factor in (\ref{k-bosons}). They satisfy the CCR:
\begin{equation}\label{CCR}
[b_k, b^*_{k^\prime}] = \delta_{k,k^\prime} \mathbb{1} , \quad
[b_k, b_{k^\prime}] =  [b^*_k, b^*_{k^\prime}] = 0 \qquad
( k,k^\prime = 0, 1, 2, \ldots, N)
\end{equation}
on the algebraic tensor product $(\mathscr{F}_{\mbox{\tiny fin}})^{\otimes (N+1)}$.

Recall that non-autonomous system with Hamiltonian for time-dependent \textit{repeated} {harmonic} perturbation
proposed in \cite{TZ} has the form
\begin{eqnarray}\label{Ham-Model}
H_{N}(t) := E b_0^*b_0 + \epsilon \sum_{k=1}^{N} b_k^*b_k +
\eta \, \sum_{k=1}^{N}\chi_{[(k-1)\tau, k\tau)}(t)\,  (b_0^*b_k +  \ b_k^*b_0) \ .
\end{eqnarray}
Here $t\in [0, N\tau)$, the parameters: $\tau, E, \epsilon, \eta$ are \textit{positive}, and $\chi_{[x,y)}(\cdot)$ is the
characteristic function of the semi-open interval $[x,y)\subset \mathbb{R}$.
It is obvious that  $H_{N}(t)$ is a self-adjoint operator with time-independent domain
\begin{equation}\label{domH}
           \mathcal{D}_0 = \bigcap_{k=0}^{N}{\rm{dom}} \, (b^*_{k}b_{k}) \subset \mathscr{H}^{(N)} \, .
\end{equation}
The model (\ref{Ham-Model}) presents the system $\mathcal{S} + \mathcal{C}_N $, where $\mathcal{S}$ is the quantum one-mode
\textit{cavity}, which is repeatedly perturbed by a time-equidistant \textit{chain} of
subsystem: $\mathcal{C}_N = \mathcal{S}_1 + \mathcal{S}_2 + \ldots + \mathcal{S}_N $.
Here $\{\mathcal{S}_k\}_{k\geq 1}$ can be considered as \textit{``atoms"} with harmonic internal degrees of freedom.
This interpretation is motivated by certain physical models known as the ``one-atom maser" \cite{BJM}, \cite{NVZ}.
The Hilbert space $\mathscr{H}_{\mathcal{S}} := \mathscr{H}_0$ corresponds to subsystem $\mathcal{S}$ and the Hilbert space
$\mathscr{H}_k$ to subsystems $\mathcal{S}_k \; $ ($k=1, \ldots, N$), respectively.
Then (\ref{H-space}) is
\begin{equation}\label{HS-HC}
\mathscr{H}^{(N)} = \mathscr{H}_{\mathcal{S}}\otimes \mathscr{H}_{\mathcal{C}_N} \, , \ \
\mathscr{H}_{\mathcal{C}_N}: = \bigotimes_{k=1}^{N} \mathscr{H}_k   \ .
\end{equation}
By (\ref{Ham-Model}) only \textit{one} subsystem $\mathcal{S}_{n}$ interacts with $\mathcal{S}$ for $t\in[(n-1)\tau, n\tau)$.
In this sense, the interaction is \textit{tuned}  \cite{TZ}.
The system $\mathcal{S}+\mathcal{C}_N$ is \textit{autonomous} on each interval
$[(n-1)\tau, n\tau)$ governed by the self-adjoint Hamiltonian
\begin{equation}\label{Ham-n}
H_n := E \, b_0^*b_0 + \epsilon\sum_{k=1}^{N}b_k^*b_k  + \eta \, (b_0^*b_n +  b_n^*b_0) \ , \ n=1,2, \ldots , N \ ,
\end{equation}
on domain $\mathcal{D}_0$.
Note that if
\begin{equation}\label{H4}
\eta^2 \leqslant E \, \epsilon \ ,
\end{equation}
Hamiltonians (\ref{Ham-Model}) and  (\ref{Ham-n}) are semi-bounded from below.
\medskip

We denote by $\mathfrak{C}_{1}(\mathscr{H}^{(N)})$ the Banach space of the \textit{trace-class}
operators on $\mathscr{H}^{(N)}$. Its dual space is isometrically isomorph to the Banach space of bounded operators on
$\mathscr{H}^{(N)}$: $\mathfrak{C}_{1}^{\ast}(\mathscr{H}^{(N)})\simeq \mathcal{L}(\mathscr{H}^{(N)})$.
The corresponding dual pair is defined by the bilinear functional
\begin{equation}\label{dual-Tr}
\langle \phi \, |  A \rangle_{\mathscr{H}^{(N)}} = \Tr_{\mathscr{H}^{(N)}} (\phi \, A)
\quad \mbox{for} \ \ (\phi, A) \in
\mathfrak{C}_{1}(\mathscr{H}^{(N)}) \times \mathcal{L}(\mathscr{H}^{(N)}) \ .
\end{equation}

The positive operators $\rho\in\mathfrak{C}_{1}(\mathscr{H}^{(N)})$ with {unit} trace is the set of \textit{density matrices}.
Recall that the state $\omega_{\rho} $ over $\mathcal{L}(\mathscr{H}^{(N)})$ is \textit{normal} if there is a density matrix
$\rho$ such that
\begin{equation}\label{state-S}
\omega_{\rho}(\, \cdot \,) = \langle \rho \, | \, \cdot \, \rangle_{\mathscr{H}^{(N)}} \ .
\end{equation}

\subsection{Master equation}
To make the system $\mathcal{S}+ \mathcal{C}_N$ open, we couple it to the \textit{boson} reservoir $\mathcal{R}$,  \cite{AJP3}.
More precisely, we follow the scheme $(\mathcal{S} + \mathcal{R}) + \mathcal{C}_N$, i.e. we study repeated perturbation of
the open system $\mathcal{S} + \mathcal{R}$ \cite{NVZ}.

Evolution of normal states of the open system $(\mathcal{S} + \mathcal{R}) + \mathcal{C}_N$ can be described by the
Kossakowski-Lindblad-Davies {dissipative} extension of the Hamiltonian dynamics to the Markovian dynamics with the
time-dependent generator \cite{AL}, \cite{AJP2}
\begin{eqnarray}\label{K-L-D-Generator}
&&L_{\sigma}(t)(\rho) := -i \, [H_N(t),\rho] + \\
&&\hspace{2.5cm} + \, \mathcal{Q}(\rho) -
\frac{1}{2} ({\mathcal{Q}}^{\ast}(\mathbb{1}) \rho + \rho \, {\mathcal{Q}}^{\ast}(\mathbb{1})) \, , \nonumber
\end{eqnarray}
for $ t\geqslant 0  $ and $\rho \in {\rm{dom}} L_{\sigma}(t)\subset \mathfrak{C}_{1}(\mathscr{H}^{(N)})$. Here the first
operator $\mathcal{Q}: \rho \mapsto \mathcal{Q}(\rho) \in \mathfrak{C}_{1}(\mathscr{H}^{(N)})$ in the dissipative part of
(\ref{K-L-D-Generator}) has the form:
\begin{equation}\label{K-L-D1}
\mathcal{Q}(\cdot) = \sigma_{-} \, b_{0} \, (\cdot) \, b^*_{0} + \sigma_{+} \, b^*_{0}\, (\cdot) \, b_{0} \ , \ \
\sigma_\mp \geqslant 0 \ ,
\end{equation}
and the operator ${\mathcal{Q}}^{\ast}$ is its dual via relation
$\langle \mathcal{Q}(\rho) \, |  A \rangle_{\mathscr{H}^{(N)}} =
\langle \rho \, | {\mathcal{Q}}^{\ast}(A) \rangle_{\mathscr{H}^{(N)}}$:
\begin{equation}\label{K-L-D2}
{\mathcal{Q}}^{\ast}(\cdot) = \sigma_{-} \, b^{*}_{0} \, (\cdot) \, b _{0} + \sigma_{+} \, b_{0}\, (\cdot) \, b^*_{0} \, .
\end{equation}
By virtue of (\ref{Ham-Model}), for $t\in[(n-1)\tau, n\tau)$, the generator (\ref{K-L-D-Generator}) takes the form
\begin{align}\label{Generator-KLD}
L_{\sigma,n}(\rho):= -i[H_n, \rho ] +  \, \mathcal{Q}(\rho) -
\frac{1}{2} ({\mathcal{Q}}^{\ast}(\mathbb{1}) \rho + \rho {\mathcal{Q}}^{\ast}(\mathbb{1})) \, .
\end{align}

The mathematical problem concerning the open quantum system is to solve the Cauchy problem for the non-autonomous
quantum Master Equation \cite{AJP2}
\begin{equation}\label{QME-sigma}
\partial_{t}\rho(t) = L_{\sigma}(t)(\rho(t)) \ , \quad \rho(0) =\rho \, .
\end{equation}
For the tuned repeated perturbation, this solution is a strongly continuous family $\{ T_{t,0}^{\sigma} \}_{t\geq 0}$,
which is defined by composition of the one-step evolution semigroups:
\[
T_{t,0}^\sigma = T_{t,(n-1)\tau}^\sigma \, T_{n-1}^\sigma \ldots T_{2}^\sigma
\ T_{1}^\sigma \ ,
\]
where $t = (n-1)\tau + \nu(t)$, $n\leqslant N, \nu(t) < \tau$. Here we put
\begin{equation}\label{One-step-T-sigma}
T_k^\sigma := T_{k}^{\sigma}(\tau) , \qquad
T_{k}^{\sigma}(s) := e^{s L_{\sigma,k}} \quad (s \geqslant 0) ,
\end{equation}
and then $ T_{t,(n-1)\tau}^\sigma = T_{n}^{\sigma}(\nu(t)) $ holds.
The evolution map is connected to solution of the Cauchy problem (\ref{QME-sigma}) by
\begin{equation}\label{T-sigma-t}
T_{t,0}^\sigma : \rho \mapsto \rho(t) = T_{t,0}^\sigma (\rho) .
\end{equation}

The construction of unique positivity- and trace-preserving dynamical semigroup on $\mathfrak{C}_{1}(\mathscr{H}^{(N)})$
for  \textit{unbounded} generator (\ref{Generator-KLD}) is a nontrivial problem. It is done in \cite{TZ1} under the
conditions (\ref{H4}) and
\begin{equation}\label{H-sigma}
            0 \leqslant \sigma_{+} < \sigma_{-} \ .
\end{equation}
for the coefficients in (\ref{K-L-D1}, \ref{K-L-D2}). Then, $\{T_{k}^{\sigma}(s)\}_{s\geqslant 0}$ for each $k$
(\ref{One-step-T-sigma}) is the Markov  dynamical semigroup, and (\ref{T-sigma-t}) is automorphism on the set of density matrices.

\subsection{Evolution in the dual space}
In order to control the evolution of normal states, it is usual to consider the $W^*$-{dynamical system}
$(\mathcal{L}(\mathscr{H}^{(N)}), \{T_{t,0}^{\sigma \;\ast}\}_{t\geqslant 0})$,
where $\{T^{\sigma \; *}_{t,0}\}_{t\geqslant 0}$ are
weak*-continuous evolution maps on the von Neumann algebra
$\mathcal{L}(\mathscr{H}^{(N)})\simeq \mathfrak{C}^*_{1}(\mathscr{H}^{(N)})$
\cite{AJP1}.
They are dual to the evolution (\ref{T-sigma-t}) on
$\mathfrak{C}_{1}(\mathscr{H}^{(N)})$ by the relation (\ref{dual-Tr}):
\begin{equation}\label{T-*sigma}
\langle T^{\sigma}_{t,0}(\rho)\;|\; A \rangle_{\mathscr{H}^{(N)}}=
\langle \rho \;|\; T^{\sigma \; *}_{t,0}(A)\rangle_{\mathscr{H}^{(N)}}  \ \quad \mbox
{for} \ \ (\rho, A) \in \mathfrak{C}_{1}(\mathscr{H}^{(N)}) \times \mathcal{L}(\mathscr{H}^{(N)}) \, ,
\end{equation}
which uniquely defines the map $A \mapsto T^{\sigma \; *}_{t,0}(A)$ for $A\in \mathcal{L}(\mathscr{H}^{(N)})$.
The corresponding dual time-dependent generator is formally given by
\begin{eqnarray}\label{K-L-D-Generator-dual}
            &&L_{\sigma}^{*}(t)(\cdot) = i \, [H_N(t),\cdot \, ] + \\
            &&\hspace{1.7cm} + \, {\mathcal{Q}}^{\ast}(\cdot) -
             \frac{1}{2} ({\mathcal{Q}}^{\ast}(\mathbb{1}) (\cdot)
            + (\cdot) {\mathcal{Q}}^{\ast}(\mathbb{1}))
             \quad \mbox{for}  \quad t\geqslant 0 \, . \nonumber
\end{eqnarray}
When $t\in[(k-1)\tau,k\tau)$, the above generator has the form
\begin{align}\label{Generator-KLD-dual}
L_{\sigma,k}^{\ast}(\cdot) = i[H_k, \cdot] +
 \, \mathcal{Q}^{\ast}(\cdot) -
\frac{1}{2} ({\mathcal{Q}}^{\ast}(\mathbb{1}) (\cdot) + (\cdot) {\mathcal{Q}}^{\ast}(\mathbb{1})) \, .
\end{align}
We adopt the notations
\begin{equation}\label{One-step-T*-sigma}
      T_k^{\sigma \, *} = T_{k}^{\sigma }(\tau)^* \ , \
    T_{t, \,(n-1)\tau}^{\sigma\; *} = T_{n}^{\sigma }(\nu(t))^* \ , \ {\rm{ and }} \ \
T_{k}^{\sigma}(s)^* := e^{s L_{\sigma,k}^*} \quad (s \geqslant 0) \ ,
\end{equation}
dual to (\ref{One-step-T-sigma}) for $t = (n-1)\tau + \nu(t)$,
$n\leqslant N$, $\nu(t) < \tau$.
Then, we obtain
\begin{align}\label{Sol-Liouv-Eq*-sigma}
         T_{t,0}^{\sigma\; *}(A) = T_{1}^{\sigma\; *}T_{2}^{\sigma\; *}\, ... \,
        T_{n-1}^{\sigma\; *} T_{t,(n-1)\tau}^{\sigma\; *}(A) \quad \mbox{ for }
        \ A \in \mathcal{L}(\mathscr{H}^{(N)}) \, .
\end{align}

\medskip

Let $\mathscr{A}(\mathscr{F})$ (or CCR($\mathbb{C}$) ) denote the Weyl
CCR-algebra on $\mathscr{F}$.
This unital $C^*$-algebra is generated as operator-norm completion of the
linear span $\mathscr{A}_w$ of the set of Weyl operators
\begin{equation}\label{S-Weyl}
     \widehat{w}(\alpha)= e^{i{\Phi(\alpha)}}  \qquad ( \, \alpha \in \C \, ),
\end{equation}
where $\Phi(\alpha) = (\overline{\alpha}a + \alpha a^*)/\sqrt{2}$ is the self-adjoint
Segal operator in $\mathscr{F}$.
[The closure of the sum is understood.]
Then CCR (\ref{CCR0}) take the Weyl form
\begin{equation}\label{W-CCR}
 \widehat{w}(\alpha_1) \widehat{w}(\alpha_2) = e^{- {i} \, {\rm{Im}}(\overline{\alpha}_{1} \alpha_{2})/2} \
 \widehat{w}(\alpha_1 + \alpha_2) \
 \qquad \mbox{for} \qquad \alpha_1,  \alpha_2 \in \C \, .
\end{equation}
We note that $\mathscr{A}(\mathscr{F})$ is contained in the $C^*$-algebra $\mathcal{L}(\mathscr{F})$
of all \textit{bounded} operators on $\mathscr{F}$.

Similarly we define the Weyl CCR-algebra $\mathscr{A}(\mathscr{H}^{(N)})
\subset \mathcal{L}(\mathscr{H}^{(N)})$ over $\mathscr{H}^{(N)} $.
This algebra is generated by operators
\begin{equation}\label{Wz}
W(\zeta) = \bigotimes_{j=0}^N \widehat{w}(\zeta_j)\ \ \ {\rm{for}} \ \
\zeta = \begin{pmatrix}   \zeta_0 \\ \zeta_1 \\  \cdot \\
           \cdot \\ \cdot \\ \zeta_N \end{pmatrix}  \in \C^{N + 1} \, .
\end{equation}
By (\ref{k-bosons}), the Weyl operators (\ref{Wz}) can be rewritten as
\begin{equation}\label{Wz-bis}
W(\zeta) = \exp[i {\big(\langle \zeta, b\rangle + \langle b, \zeta\rangle\big)}/\sqrt 2] \ ,
\end{equation}
where the sesquilinear form notations
\begin{equation}\label{b-forms}
      \langle \zeta, b\rangle  := \sum_{j=0}^N \bar{\zeta}_j b_j,  \qquad
      \langle b, \zeta\rangle  := \sum_{j=0}^N  \zeta_j b^*_j
\end{equation}
are used.
Let us recall that $\mathscr{A} (\mathscr{H}^{(N)}) $ is weakly dense
in $\mathcal{L}(\mathscr{H}^{(N)})$\cite{AJP1}.

\smallskip

Explicit formulae for evolution operators (\ref{One-step-T*-sigma}) acting on the Weyl operators has been established
in \cite{TZ1}. For $ n=1, 2, \ldots. N$, let $J_n$ and $X_n$ be $(N+1)\times (N+1)$ Hermitian matrices:
\begin{equation}\label{J-n}
      (J_n)_{jk} = \begin{cases}
                      1 & \quad ( j=k=0 \; \mbox{ or } \; j=k=n) \\
                      0 & \quad \mbox{otherwise}
                   \end{cases} ,
\end{equation}
\begin{equation}\label{X-n}
     (X_n)_{jk} = \begin{cases}
                      (E-\epsilon)/2 & \quad (j,k)=(0,0)  \\
                      -(E-\epsilon)/2 & \quad (j,k)=(n,n)  \\
                      \eta & \quad (j,k)=(0,n) \\
                      \eta & \quad (j,k)=(n,0) \\
                      0 & \quad \mbox{otherwise}
                   \end{cases} .
\end{equation}
We define the matrices
\begin{equation}\label{Y-n}
    Y_n: =\epsilon I + \frac{E-\epsilon}{2}J_n + X_n \  \ (n= 1, \ldots ,N) \, ,
\end{equation}
where $I$ is the $(N+1)\times (N+1)$ identity matrix. Then Hamiltonian (\ref{Ham-n}) takes the form
\begin{equation}\label{HY}
H_n={\sum_{j,k=0}^{N} (Y_n)_{jk}b_j^*b_k} .
\end{equation}
We also need the $(N+1)\times (N+1)$ matrix $P_0$ defined by
$(P_0)_{j k} = \delta_{j 0}\delta_{k 0} \,$ $(j, k = 0, 1, 2, \ldots, N)$.
Then one obtains the following proposition which is proved in \cite{TZ1}:
\begin{prop}\label{one-step-dualT}
Let $n = 1, 2, \ldots, N $ and $\zeta \in \C^{N + 1}$.
Then for  $s\geqslant 0$, the dual Markov dynamical semigroup (\ref{One-step-T*-sigma}) on the Weyl $C^*$-algebra
has the form
\begin{equation}\label{T*-n}
{T}_{n}^{\sigma *}(s)(W(\zeta))  = \Omega_{n,s}^{\sigma}(\zeta) W(U_n^{\sigma}(s)\zeta) \ ,
\end{equation}
where
\begin{equation}\label{Omega}
\Omega_{n,s}^{\sigma}(\zeta) :=
\exp \! \Big[ \, - \frac{1}{4} \ \frac{\sigma_- + \sigma_+}{\sigma_- - \sigma_+}
\big(\langle \zeta, \zeta \rangle  - \langle U_n^{\sigma}(s)\zeta, U_n^{\sigma}(s)\zeta \rangle\big) \Big]
\end{equation}
and
\begin{equation}\label{U-n-sigma}
U_n^{\sigma}(s) = \exp \! \Big[ \, i \, s \Big(Y_n  + i \ \frac{\sigma_- - \sigma_+}{2}P_0\Big)\Big]
\end{equation}
under the conditions (\ref{H4}) and (\ref{H-sigma}).
Therefore, the $k$-step evolution ($t=k\tau , k\leqslant N$ in (\ref{Sol-Liouv-Eq*-sigma})) of the Weyl operator  is given by
\begin{equation*}\label{T*-k}
T_{k\tau,0}^{\sigma \; \; *}(W(\zeta)) = \exp \! \Big[\, - \frac{\sigma_- + \sigma_+}{4(\sigma_- - \sigma_+)}
\big(\langle \zeta, \zeta \rangle -
\langle U_1^{\sigma} \ldots U_k^{\sigma} \, \zeta, \, U_1^{\sigma} \ldots U_k^{\sigma} \, \zeta \rangle \big) \Big]
\end{equation*}
\begin{equation}\label{T*-k-W}
              \times \   W(U_1^{\sigma} \ldots U_k^{\sigma} \, \zeta) \ ,
\end{equation}
where $T_{k\tau,0}^{\sigma \; \; *}= T_{1}^{\sigma\; *}T_{2}^{\sigma\; *} \, \ldots \, T_{k}^{\sigma\; *} $
and $U_n^{\sigma} := U_n^{\sigma}(\tau)$.
\end{prop}
\begin{remark}\label{z-w-g-sigma}
The explicit expression of the matrix $U_n^{\sigma}(t)$ in  (\ref{U-n-sigma}) is given by
$U_n^{\sigma}(t) = e^{it\epsilon}V_n^{\sigma}(t)$, where
\begin{equation}\label{V-s}
       (V^{\sigma}_n(t))_{jk} = \begin{cases}
        g^{\sigma}(t)z^{\sigma}(t)\,\delta_{k0} +
       g^{\sigma}(t)w^{\sigma}(t)\,\delta_{kn} & \quad (j = 0) \\
    g^{\sigma}(t){w^{\sigma}(t)}\,\delta_{k0}
       + g^{\sigma}(t){z^{\sigma}(-t)}\,\delta_{kn} &
        \quad (j = n) \\
                   \qquad \delta_{jk}    & \quad (\mbox{otherwise})
                    \end{cases} .
\end{equation}
Here $E_{\sigma } := E + i \ (\sigma_- - \sigma_+)/{2}$ and
\begin{equation}\label{g-w-s}
      g^{\sigma}(t) := e^{it(E_{\sigma}-\epsilon)/2}, \qquad
           w^{\sigma}(t) :=\frac{2i\eta}{\sqrt{(E_{\sigma}-\epsilon)^2+4\eta^2}}
          \sin t\sqrt{\frac{(E_{\sigma}-\epsilon)^2}{4}+\eta^2} \ ,
\end{equation}
\begin{equation}\label{z-s}
          z^{\sigma}(t) := \cos t\sqrt{\frac{(E_{\sigma}-\epsilon)^2}{4}+\eta^2}
          +\frac{i(E_{\sigma}-\epsilon)}{\sqrt{(E_{\sigma}-\epsilon)^2+4\eta^2}}
          \sin t\sqrt{\frac{(E_{\sigma}-\epsilon)^2}{4}+\eta^2} \ .
\end{equation}
Note that the relation  $ z^{\sigma}(t)z^{\sigma}(-t) - w^{\sigma}(t)^2  = 1 $  holds
for any $\sigma_{\pm} \geqslant 0$,
whereas one has $|g^{\sigma}(t)|^2(|z^{\sigma}(t)|^2 + |w^{\sigma}(t)|^2) < 1$ and
$z^{\sigma}(-t) \ne \overline{z^{\sigma}(t)}$ for $ 0 \leqslant \sigma_{+} < \sigma_{-}$.

Hereafter, together with (\ref{T*-k-W}) we also use the following short-hand notations:
\begin{equation}\label{g-w-z-s}
g^{\sigma} = g^{\sigma}(\tau)  , \  w^{\sigma} :=w^{\sigma}(\tau) , \
z^{\sigma} = z^{\sigma}(\tau) \ \
{\rm{and}} \ \ V_n^{\sigma} :=V_n^{\sigma}(\tau) \, .
\end{equation}
\end{remark}
\begin{remark}\label{quasi-free}
Dual dynamical semigroups (\ref{T*-n}) and the evolution operator (\ref{T*-k-W}) are examples of the quasi-free maps
on the Weyl $C^*$-algebra. Using the arguments of \cite{DVV}, we have shown in
\cite{TZ1} that they can be extended to the unity-preserving completely positive
linear maps on $\mathcal{L}(\mathscr{H}^{(N)})$ under the conditions (\ref{H4})
and (\ref{H-sigma}).
\end{remark}

\smallskip

The aim of the rest of the paper is to study evolution of the reduced density
matrices for subsystems of the total system $(\mathcal{S} + \mathcal{R})+\mathcal{C}_N$.

In Section \ref{RDM-I}, we consider the subsystem $\mathcal{S}$.
This includes analysis of convergence to stationary states in the infinite-time limit $N\rightarrow \infty$.
We also perform a similar analysis for the subsystems $\mathcal{S} + \mathcal{S}_m$ and $\mathcal{S}_m + \mathcal{S}_n$.
Section \ref{RDM-subS} is devoted to a more complicated problem of evolution of reduced density matrices for finite
subsystems, which include $\mathcal{S}$ and a part of $\mathcal{C}_N$.
This allows us to detect an asymptotic behaviour of the quantum correlations between
$\mathcal{S}$ and a part of $\mathcal{C}_N$ caused by repeated perturbation and dissipation
for large $N$ in terms of those for small $N$ with the stable initial state.

For the brevity, we hereafter supress the dependence on $N $ of the Hilbert space
$\mathscr{H}^{(N)} $ as well as of the Hamiltonian $H_N(t) $ and the subsystem
$\mathcal{C}_N $, when it will not cause any confusion.
\section{Time Evolution of Subsystems I} \label{RDM-I}
\subsection {Subsystem \textit{S}} \label{preliminaries}
We start by analysis of the simplest subsystem $\mathcal{S}$. Let the initial state of the total system
$\mathcal{S} + \mathcal{C}$ be defined by a density matrix
$\rho  \in {\mathfrak{C}}_{1}(\mathscr{H}_{\mathcal{S}} \otimes \mathscr{H}_{\mathcal{C}})$.
Then for any $t\geqslant 0$, the evolved state $\omega_{\mathcal{S}}^{t}(\cdot)$ on the Weyl $C^*$-algebra
$\mathscr{A}(\mathscr{H}_{\mathcal{S}})$ of subsystem $\mathcal{S}$ is given by the partial trace:
\begin{equation}\label{st-S}
     {\omega_{\mathcal{S}}^{t}(A) = \omega_{\rho(t)}(A \otimes \mathbb{1})} =
      \Tr_{\mathscr{H}_{\mathcal{S}} \otimes \mathscr{H}_{\mathcal{C}_N}}
(T_{t,0}^{\sigma}(\rho_\mathcal{S} \otimes \rho_\mathcal{C})\, A\otimes\mathbb{1})
\quad \mbox{for} \quad  A \in  \mathscr{A}(\mathscr{H}_{\mathcal{S}}) \ ,
\end{equation}
where  $\rho(t) = T_{t,0}^{\sigma}\rho$ and $ \mathbb{1} \in \mathscr{A}(\mathscr{H}_{\mathcal{C}}) $.
Recall that for a density matrix $\varrho \in \mathfrak{C}_{1}(\mathscr{H}_{\mathcal{S}}\otimes \mathscr{H}_{\mathcal{C}})$,
the \textit{partial trace} of $\varrho$ with respect to the Hilbert space $\mathscr{H}_\mathcal{C}$ is a bounded linear
map $\Tr_{\mathscr{H}_{\mathcal{C}}}: \varrho \mapsto \widehat{\varrho} \in \mathfrak{C}_{1}(\mathscr{H}_{\mathcal{S}})$
characterised by the identity
\begin{equation}\label{P-Tr}
\Tr_{\mathscr{H}_{\mathcal{S}} \otimes \mathscr{H}_{\mathcal{C}}} (\varrho \,  (A \otimes \mathbb{1})) =
\Tr_{\mathscr{H}_{\mathcal{S}}} (\widehat{\varrho} \, A ) \quad {\rm for}  \quad  A \in  \mathcal{L}(\mathscr{H}_{\mathcal{S}})\ .
\end{equation}
If one puts
\begin{equation}\label{P-Tr-t}
       \rho_{\mathcal{S}}(t) := \Tr_{\mathscr{H}_{\mathcal{C}}}
       (T_{t,0}^{\sigma} (\rho)) \, ,
\end{equation}
then one gets  the identity
\begin{equation}\label{st-S-matrix}
     \omega_{\mathcal{S}}^{t}(A) =
         \Tr_{\mathscr{H}_{\mathcal{S}}} (\rho_{\mathcal{S}}(t) \, A ) =: \omega_{\rho_{\mathcal{S}}(t)}(A) \ ,
\end{equation}
by (\ref{st-S}), i.e., $ \rho_{\mathcal{S}}(t) $ is the density matrix defining the normal state $\omega^t_{\mathcal{S}}$.

In the followings, we mainly consider the \textit{initial} density matrices of the form:
\begin{equation}
\rho = \rho_{\mathcal{S}} \otimes \rho_{\mathcal{C}} \quad {\rm{for}} \quad
\rho_{\mathcal{S}} = \rho_0 \ , \quad \rho_{\mathcal{C}}
= \bigotimes_{k=1}^N \rho_k   \quad\mbox{with}\quad \rho_1= \rho_2 = \ldots = \rho_N  \, .
\label{state-S-C1}
\end{equation}
Note that the \textit{characteristic function} $E_{\omega_\mathcal{S}}\!\!:\!\C\!\to\!\C$ of the state $\omega_\mathcal{S}$ on
the algebra $\mathscr{A}(\mathscr{H}_S)$ is
\begin{equation}\label{E-state-S}
E_{\omega_\mathcal{S}}(\theta) = \omega_\mathcal{S} (\widehat w(\theta))
\end{equation}
and that (\ref{E-state-S}) can uniquely determine the state $\omega_\mathcal{S}$ by the Araki-Segal theorem \cite{AJP1}.
\begin{lem} \label{lem-1-step-evol} Let $A = \widehat w(\theta)$. Then evolution of (\ref{st-S}) on the interval $[ 0, \tau)$
yields
\begin{equation*}
E_{\omega_{\mathcal{S}}^{t}}(\theta) =
  \exp\Big[-\frac{|\theta|^2}{4}\frac{\sigma_- +\sigma_+}{\sigma_- - \sigma_+}
        \big(1 - |g^{\sigma}(t) z^{\sigma}(t)|^2
      - |g^{\sigma}(t)  w^{\sigma}(t)|^2 \big)\Big]
\end{equation*}
\begin{equation}\label{one-step-evol1}
             \times \omega_{\rho_0}\big(\widehat w( e^{i\tau\epsilon}
          g^{\sigma}(t) z^{\sigma}(t) \theta ) \big)
      \omega_{\rho_1}\big(\widehat w(e^{i\tau\epsilon}  g^{\sigma}(t)
      w^{\sigma}(t) \theta) \big) \  , \   t\in[0, \tau) \ .
\end{equation}
\end{lem}
{\sl Proof :} By (\ref{Wz}), we obtain that $W(\theta e) = \widehat w(\theta) \otimes
\mathbb{1}\otimes \ldots \otimes \mathbb{1}$ for the vector
$ \,  e = {}^t( 1,  0,  \ldots,  0)  \in \C^{N+1} \, $,
where ${{}^t(\ldots)}$ means the \textit{vector-transposition}, cf (\ref{Wz}).
Then (\ref{st-S})-(\ref{st-S-matrix}) yield
\begin{equation}\label{initial-1step}
\omega_{\mathcal{S}}^{t}(\widehat w(\theta))= \omega_{\rho(t)}
(\widehat w(\theta)\otimes \mathbb{1}\otimes \ldots \otimes \mathbb{1}) =
\omega_{\rho_{\mathcal{S}}(t)}(\widehat w(\theta))   \, .
\end{equation}
By virtue of duality (\ref{T-*sigma}) and (\ref{T*-k-W}) for $k=1$,  we obtain
\[
\omega_{\rho_{\mathcal{S}}(t)}(\widehat w(\theta))=
\omega_{\rho_{\mathcal{S}} \otimes \rho_{\mathcal{C}}}((T_{t,0}^{\sigma \; *}W)(\theta e)) =
\omega_{\bigotimes_{j=0}^N \rho_j }(({T}_{t,0}^{\sigma \, *} W)(\theta e))
\]
\[      = \exp\Big[-\frac{|\theta|^2}{4}\frac{\sigma_- +\sigma_+}{\sigma_- - \sigma_+}
         \big(1- \langle U_1^{\sigma}(t) e, U_1^{\sigma}(t) e \rangle
     \big) \Big] \omega_{\bigotimes_{j=0}^N \rho_j }
      \big(W(\theta \, U^{\sigma}_1(t)e)\big) \, .
\]
Taking into account (\ref{V-s}) and (\ref{E-state-S}), one obtains for (\ref{initial-1step}) the expression
which coincides with assertion (\ref{one-step-evol1}).
\hfill $\square$
\medskip

Similarly, for $t= m \tau$  we obtain  the characteristic function
\[
E_{\omega_{\mathcal{S}}^{m \tau}}(\theta)=
\omega_{\rho_{\mathcal{S}} \otimes \rho_{\mathcal{C}}}
         (T_{m\tau,0}^{\sigma \, *}(W(\theta  e))
         = \exp\Big[-\frac{|\theta|^2}{4}\frac{\sigma_- +\sigma_+}{\sigma_- - \sigma_+}
         \big(1- \langle U_1^{\sigma}\ldots U^{\sigma}_m e,
       U_1^{\sigma}\ldots U^{\sigma}_m e \rangle
     \big) \Big]
\]
\begin{equation}\label{m-step-evol}
       \times \omega_{\bigotimes_{j=0}^N \rho_j }
      \big(W(\theta \, U^{\sigma}_1 \ldots U^{\sigma}_m e)\big) = \,
\end{equation}
\[
= \exp\Big[-\frac{|\theta|^2}{4}\frac{\sigma_- +\sigma_+}{\sigma_- - \sigma_+}
         \big(1- \langle U_1^{\sigma}\ldots U^{\sigma}_m e,
       U_1^{\sigma}\ldots U^{\sigma}_m e \rangle
     \big) \Big]
   \prod_{j=0}^N\omega_{\rho_j } \big(
     \widehat w(\theta \, (U^{\sigma}_1 \ldots U^{\sigma}_m e)_j)\big) \, ,
\]
where we have used (\ref{Wz}) and (\ref{T*-k-W}). By (\ref{V-s}) we obtain
{\begin{equation}\label{0Uk0}
(U_1^{\sigma} \ldots U_m^{\sigma} \  e)_{k} =  \begin{cases}
          e^{iN\tau\epsilon} (g^{\sigma}(\tau)z^{\sigma}(\tau))^m   \ & \ (k=0)  \\
          e^{iN\tau\epsilon}g^{\sigma}(\tau) w^{\sigma}(\tau) (g^{\sigma}(\tau)z^{\sigma}(\tau))^{m-k}  \ & \
          (1\leqslant k \leqslant m) \\
           0  \ & \  ( m < k \leqslant N )   \, .
\end{cases}
\end{equation}}
Then taking into account $|g^{\sigma}z^{\sigma}| < 1$ (Remark \ref{z-w-g-sigma}), we find
\begin{eqnarray}\label{scal-prod}
&&\langle  e,  e \rangle -
\langle U_1^{\sigma} \ldots U_m^{\sigma} \,  e \, ,
U_1^{\sigma} \ldots U_m^{\sigma} \,  e \rangle  \\
&& = (1 - |g^{\sigma}z^{\sigma}|^{2m})\Big[1 - \frac{|g^{\sigma} w^{\sigma}|^2}
{1 - |g^{\sigma}z^{\sigma}|^2 }  \Big] \, . \nonumber
\end{eqnarray}
By setting $m=N$, (\ref{E-state-S}), (\ref{m-step-evol})-(\ref{scal-prod}) yield
the following result.
\begin{lem} \label{corr-m-step-evol} The state of the subsystem $\mathcal{S}$ after $N$-step evolution has the
characteristic function
\begin{eqnarray}\label{m-step-evol-bis}
&& E_{\omega_{\mathcal{S}}^{N \tau}}(\theta) =  \omega_{\rho_{\mathcal{S}}(N\tau)}(\widehat w(\theta))  \\ \nonumber
&& = \exp\Big[-\frac{|\theta|^2}{4}\frac{\sigma_- +\sigma_+}{\sigma_- - \sigma_+}
(1 - |g^{\sigma}z^{\sigma}|^{2N})\Big(1 - \frac{|g^{\sigma} w^{\sigma}|^2}
{1 - |g^{\sigma}z^{\sigma}|^2 } \Big)\Big]\\  \nonumber
&&\times \omega_{\rho_0}\big(\widehat w(e^{iN\tau\epsilon} (g^{\sigma})^{N} (z^{\sigma})^{N} \theta) \big)
\prod_{k=1}^{N}\omega_{\rho_k}\big(\widehat w( e^{iN\tau\epsilon}
(g^{\sigma})^{N-k+1} (z^{\sigma})^{N-k} w^{\sigma} \theta) \big) \, . \nonumber
\end{eqnarray}
\end{lem}

To consider the asymptotic behaviour of the state $\omega_{\mathcal{S}}^{N\tau}$ for large  $N$,
we assume that the state $\omega_{\rho_k}$
on $\mathscr{A}(\mathscr{F})$ is \textit{gauge-invariant}, i.e.,
\begin{equation}\label{g-inv}
e^{- i \,\phi \, a^* a} \rho_k \, e^{ i \,\phi \, a^* a}
= \rho_{k}
\qquad (\phi\in \mathbb{R})
\end{equation}
for each component of the initial density matrix $\rho_{\mathcal{C}}$ (\ref{state-S-C1}).
\vspace{-0.1cm}
\begin{thm}\label{limit-S} Let $\omega_{\rho_k}$ be \textit{gauge-invariant} for $k=1,2, \ldots, N$ and suppose that the
product
\begin{equation}\label{D}
      D(\theta):= \prod_{s=0}^{\infty} \omega_{\rho_1}(\widehat w((g^{\sigma}z^{\sigma})^s \theta) )  \, ,
\end{equation}
converges for any $\theta \in \C$ and let the map $ \; \R\ni r \mapsto D(r \, \theta) \in \C \; $ be continuous.
Then for any initial normal state $\omega_{\mathcal{S}}^{0}(\cdot) = \omega_{\rho_0}(\cdot)$ of the subsystem
$\mathcal{S}$, the following properties hold. \\
(a) The pointwise limit of the characteristic functions (\ref{m-step-evol-bis}) exists
\begin{equation}\label{lim-m}
E_{\ast}(\theta)= \lim_{N \rightarrow \infty} \omega_{\rho_{\mathcal{S}}(N\tau)}(\widehat w(\theta)) \, ,
\quad  \theta \in \C \, .
\end{equation}
(b) There exists a unique density matrix $\rho_{*}^{\mathcal{S}}$  such that the limit (\ref{lim-m}) is a characteristic
function of the gauge-invariant normal state: $E_{\ast}(\theta) = \omega_{\rho_{*}^{\mathcal{S}}}(\widehat w(\theta))$. \\
(c) The states $\{\omega_{\mathcal{S}}^{m\tau}\}_{m\geqslant 1}$ converge  to
$\omega_{\rho_{*}^{\mathcal{S}}} $ for $m \rightarrow \infty$ in the weak*-topology.
\end{thm}
\vspace{-0.1cm}
{\sl Proof}: (a) By (\ref{S-Weyl}) and by the gauge-invariance (\ref{g-inv}), one gets
$\omega_{\rho_k}(\widehat w( e^{i \phi} \theta)) = \omega_{\rho_k}(\widehat w(\theta))$ for every $\phi\in \mathbb{R}$.
Hence, for $1\leqslant k\leqslant N$ the characteristic functions $E_{\omega_{\rho_k}}(\theta)$ depend only on $|\theta|$,
and we can skip the factor $e^{iN\tau\epsilon}$ in the arguments of the factors in the right-hand side of
(\ref{m-step-evol-bis}). Note that for $N \rightarrow \infty$ the factor $\omega_{\rho_0}$ converges to one,
since the normal states are regular and $|g^{\sigma}z^{\sigma}|<1$ (see Remark \ref{z-w-g-sigma}). Hence,
the pointwise limit (\ref{lim-m}) follows from (\ref{m-step-evol-bis}) and the hypothesis (\ref{D}). It does not depend on
the initial state $\omega_{\rho_0}$ of the subsystem $\mathcal{S}$ and the explicit expression of (\ref{lim-m}) is given by
\begin{equation}\label{lim-m-expl}
 E_{\ast}(\theta) =
\exp\Big[-\frac{|\theta|^2}{4} \frac{\sigma_- +\sigma_+}{\sigma_- - \sigma_+}
\Big(1 -\frac{|g^{\sigma}w^{\sigma}|^{2}} {1 - |g^{\sigma}z^{\sigma}|^2 } \Big)\Big]
D(g^{\sigma}w^{\sigma} \theta) \, .
\end{equation}
(b) The limit (\ref{lim-m-expl}) inherits the properties of characteristic functions
$E_{\omega_{\mathcal{S}}^{m\tau}}(\theta)= \omega_{\mathcal{S}}^{m\tau}(\widehat w(\theta))$:\\
(\textit{i}) normalisation: $E_{\ast}(0) = 1$ ,\\
(\textit{ii}) unitary : $\overline{E_{\ast}(\theta)} = E_{\ast}(- \theta)$ ,\\
(\textit{iii}) positive definiteness:
$\sum_{k,k' = 1}^K \, \overline{z_{k}} {z_{k'}} e^{- i \, Im(\overline{\theta_k} \theta_{k'})/2}
\, E_{\ast}(\theta_k - \theta_{k'}) \geqslant 0$ for any $K \geqslant  1$ and $z_{k} \in \mathbb{C}$
 $(k=1,2, \ldots , K) $ , \\
(\textit{iv}) regularity: the continuity of the map $r \mapsto D(r\theta)$ implies that the function
$r \mapsto E_{\ast}(r \, \theta)$ is also continuous.\\
Then by the Araki-Segal theorem, the properties (i)-(iv) guarantee the existence of the unique normal state
$\omega_{\rho_{*}^{\mathcal{S}}}$ over the CCR algebra $\mathscr{A}(\mathscr{H}_{\mathcal{S}})$ such that
$E_{\ast}(\theta) = \omega_{\rho_{*}^{\mathcal{S}}}(\widehat w(\theta))$. Taking into account (a) and (\ref{lim-m-expl})
we conclude that in contrast to the initial state $\omega_{\mathcal{S}}^{0}$ the limit state $\omega_{\rho_{*}^{\mathcal{S}}}$
is gauge-invariant.\\
(c) The convergence  (\ref{D}) can be extended by linearity to the algebraic span
of the set of Weyl operators $\{\widehat w(\alpha)\}_{\alpha \in \mathbb{C}}$.
Since it is norm-dense in $C^*$-algebra $\mathscr{A}(\mathscr{H}_{\mathcal{S}})$, the weak*-convergence of the states
$\omega_{\mathcal{S}}^{m\tau}$ to the limit state $\omega_{\rho_{*}^{\mathcal{S}}}$  follows (see \cite{BR1}, \cite{AJP1}).
\hfill $\square$
\smallskip
\begin{remark}\label{Ex} (a) By Theorem \ref{limit-S} (a)-(b), one has $\rho_{*}^{\mathcal{S}} = \rho_{*}^{\mathcal{S}}(\tau)$,
i.e. the limit state $\omega_{\rho_{*}^{\mathcal{S}}}$ is invariant under the one-step evolution $T_{\tau,0}^{\sigma}$.
Comparing (\ref{one-step-evol1}) and (\ref{lim-m-expl}) one finds that
$\rho_{*}^{\mathcal{S}} \neq \rho_{*}^{\mathcal{S}}(\nu)$ for $0< \nu <\tau$. Instead, the evolution for
repeated perturbation  yields the asymptotic periodicity:
\begin{equation}\label{asym-per}
\lim_{n \rightarrow \infty} (\omega_{\rho_{*}^{\mathcal{S}}(t)}(\widehat w(\theta)) -
\omega_{\rho_{*}^{\mathcal{S}}(\nu(t))}(\widehat w(\theta)) = 0 \ , \  {\rm{for}} \  t = (n-1)\tau + \nu(t) \ .
\end{equation}
\noindent (b) Let $\rho_1$ in (\ref{state-S-C1}) correspond to the \textit{quasi-free} gauge-invariant Gibbs state for the
inverse temperature $\beta > 0$ and let $\omega_{\rho_0}(\cdot)$ be any initial normal state of the subsystem
$\mathcal{S}$. Since
\begin{equation}\label{Gibbs-E1}
           \omega_{\rho_1}(\widehat w(\theta)) =
         \exp\Big[ - \frac{1}{4}\ |\theta|^2 \coth
      \frac{\beta}{2} \, \Big]
\end{equation}
holds, we obtain for (\ref{D}):
\begin{equation}\label{D-Gibbs}
          D(\theta)= \exp \Big[ - \frac{1}{4}\
         \frac{|\theta|^2}{1- |g^{\sigma}z^{\sigma}|^2} \
        \coth \frac{\beta}{2} \, \Big] \ .
\end{equation}
Put $ \lambda^{\sigma}(\tau) :=
    |g^{\sigma}w^{\sigma}|^{2}
   (1 - |g^{\sigma}z^{\sigma}|^2)^{-1} \in [0, 1)$ (Remark \ref{z-w-g-sigma}).
Then for the characteristic function of the limit state in Theorem \ref{limit-S}, we get
\begin{equation}\label{Gibbs-E2}
   \omega_{\rho_*}(\widehat w(\theta)) =
     \exp\left[-\frac{|\theta|^2}{4} \left((1 -\lambda^{\sigma}(\tau))
        \frac{\sigma_- +\sigma_+}{\sigma_- - \sigma_+}
         + \lambda^{\sigma}(\tau) \ \coth \frac{\beta}{2}\right)\right] .
\end{equation}

If $w^{\sigma} = 0$ (i.e. $\lambda^{\sigma}(\tau) = 0$), the subsystem $\mathcal{S}$
seems to interact only with reservoir $\mathcal{R}$,
and it evolves to a \textit{steady} state with characteristic function
\begin{equation}\label{Gibbs-E2-0}
           E_{\ast 0}(\theta) = \exp\left[-\frac{|\theta|^2}{4}
          \frac{\sigma_- +\sigma_+}{\sigma_- - \sigma_+}\right] \, , \  \  \  \  0 \leq \sigma_{+} < \sigma_{-}  \  ,
\end{equation}
which corresponds to the quasi-free Gibbs state for the inverse temperature $\beta_{\ast 0} : = \ln (\sigma_{-}/ \sigma_+)$.
This reflects thermal \textit{equilibrium} between $\mathcal{S}$ and $\mathcal{R}$.
In this sense, $\beta_{\ast 0}$ is the inverse temperature of the external reservoir $\mathcal{R}$ \cite{NVZ}.

If  $w^{\sigma} \neq 0$, the steady state (\ref{Gibbs-E2}) of subsystem
$\mathcal{S}$ has the characteristic function
\begin{equation}\label{Gibbs-E3}
E_{\ast}(\theta) = \exp \left[-\frac{|\theta|^2}{4} \
       \coth \frac{\beta_{\ast}^{\sigma}(\tau)}{2}\right] ,
\end{equation}
where the inverse temperature $\beta_{\ast}^{\sigma}(\tau)$ is defined by
\begin{equation*}
         \coth \frac{\beta_{\ast}^{\sigma}(\tau)}{2}
            = (1 -\lambda^{\sigma}(\tau))\ \coth \frac{\beta_{\ast 0}}{2}
             + \lambda^{\sigma}(\tau) \ \coth \frac{\beta}{2} \  .
\end{equation*}
Note that $\beta_{\ast}^{\sigma}(\tau)$ satisfies either
$\beta_{\ast 0} \leqslant \beta_{\ast}^{\sigma}(\tau) \leqslant \beta \, $  or
$ \, \beta_{\ast 0} \geqslant \beta_{\ast}^{\sigma}(\tau) \geqslant \beta$.
\end{remark}

\subsection {Correlations: subsystems \textit{S} + \textit{S\small{n}} and \textit{S\small{m}} + \textit{S\small{n}}}\label{S-Sm}
To study quantum correlations induced by repeated perturbation, we cast the
first glance on the \textit{bipartite} subsystems $\mathcal{S} + \mathcal{S}_n$
and $\mathcal{S}_{m} + \mathcal{S}_{n}$.
We consider the initial density matrix (\ref{state-S-C1})
satisfying
\begin{equation}\label{Gibbs}
         \omega_{\rho_0}(\widehat w(\theta)) = \exp\Big[- \frac{|\theta|^2}{4}
           \coth\frac{\beta_0}{2}\Big] \ , \
\omega_{\rho_j}(\widehat w(\theta)) = \exp\Big[- \frac{|\theta|^2}{4} \coth\frac{\beta}{2}\Big] \, .
\end{equation}
From  (\ref{T-*sigma}) and (\ref{T*-k-W}), we have:
\begin{prop}\label{Gibbs-evolution}
For evolved density matrix $\rho(N\tau) = T_{N\tau,0}^{\sigma}\, \rho$ the characteristic function of the state
$\omega_{\rho(N\tau)}(\cdot)$ is
\begin{equation}\label{G-evol}
  \omega_{\rho(N\tau)}(W(\zeta))
 = \langle \rho \, | \, T_{N\tau,0}^{\sigma \, *}(W(\zeta)) \rangle_{\mathscr{H}}
 = \exp\Big[-\frac{1}{4}\langle \zeta, X^{\sigma}(N\tau) \zeta\rangle\Big],
\end{equation}
where $ X^{\sigma}(N\tau) $ is the $(N+1)\times(N+1)$ matrix given by
\begin{eqnarray}\label{X-sigma}
 X^{\sigma}(N\tau) &=& U_N^{\sigma \; *} \ldots U_1^{\sigma \; *}
\Big[\Big( - \frac{\sigma_- + \sigma_+}{\sigma_- - \sigma_+} +
\frac{1+e^{-\beta}}{1-e^{-\beta}} \Big) I +
\Big(\frac{1+e^{-\beta_0}}{1-e^{-\beta_0}}-\frac{1+e^{-\beta}}{1-e^{-\beta}}\Big)P_0\Big]
\nonumber
\\
      && \times U_1^{\sigma} \ldots U_N^{\sigma}
      \quad + \ \frac{\sigma_- + \sigma_+}{\sigma_- - \sigma_+} I .
\end{eqnarray}
\end{prop}
\begin{remark}\label{corr-entangl}
In the theory of quantum correlation and entanglement for quasi-free states the matrix $X^{\sigma}(t)$ is known as the
\textit{covariant matrix} for Gaussian states, see  \cite{AdIl}, \cite{Ke}. Indeed, differentiating (\ref{G-evol})
with respect to components of $\zeta$ and $\overline{\zeta}$ at $\zeta =0$, one can identify the entries of
$X^{\sigma}(t)$ with expectations of monomials generated by the creation and the annihilation operators involved in
(\ref{Wz-bis}), (\ref{b-forms}).
\end{remark}

\smallskip

\noindent
\textit{Subsystem} $\mathcal{S} + \mathcal{S}_n$. For $1 < n \leqslant N$ the initial
state $\omega_{\mathcal{S} + \mathcal{S}_n}^{0}(\cdot) $ on the Weyl $C^*$-algebra
$\mathscr{A}(\mathscr{H}_{0}\otimes \mathscr{H}_{n})\simeq
\mathscr{A}(\mathscr{H}_{0})\otimes \mathscr{A}(\mathscr{H}_{n}) $
of this \textit{composed} subsystem is given by the partial trace
\begin{eqnarray}\nonumber
&&\omega_{\mathcal{S} + \mathcal{S}_n}^{0}(\widehat{w}(\alpha_0)\otimes\widehat{w}(\alpha_1)) =
\omega_{\rho}(\widehat{w}(\alpha_0)\otimes\bigotimes_{k=1}^{n-1}\mathbb{1}\otimes\widehat{w}(\alpha_1)
\otimes \bigotimes_{k=n+1}^{N}\mathbb{1})\\
&& = \exp\Big[- \frac{|\alpha_0|^2}{4} \coth\frac{\beta_{0}}{2} \Big]
\exp\Big[- \frac{|\alpha_1|^2}{4} \coth\frac{\beta}{2} \Big]\ . \label{st-S0-Sm}
\end{eqnarray}
This is the characteristic function of the product state corresponding to two isolated systems with different temperatures.
Put $\zeta^{(0,n)} := {{}^t(\alpha_0, 0, \ldots , 0, \alpha_1, 0, \ldots , 0) \in \C^{N + 1}} $,
where {$\alpha_1$} occupies the $(n+1)$th position. .
Then we get
\begin{equation}\label{omOM}
                  \omega_{\mathcal{S} +\mathcal{S}_n}^{N\tau}
        (\widehat{w}(\alpha_0)\otimes\widehat{w}(\alpha_1)) =
            \omega_{\rho(N\tau)}(W(\zeta^{(0,n)})) \ .
\end{equation}
For the components of the vector $U_1^{\sigma}\ldots U_N^{\sigma} \zeta^{(0,n)}$,  we get from Remark \ref{z-w-g-sigma} that
\begin{equation}\label{0mUk0}
(U_1^{\sigma}\ldots U_N^{\sigma} \ \zeta^{(0,n)})_k
=
\end{equation}
\[
\begin{cases}  e^{iN\tau\epsilon} \ [ ( g^{\sigma} z^{\sigma})^N \ \alpha_0 +
(g^{\sigma} z^{\sigma} )^{n-1} g^{\sigma} w^{\sigma} \, \alpha_1 ] ,
\ & (k=0) \\
e^{iN\tau\epsilon} [ (g^{\sigma} z^{\sigma} )^{N-k} g^{\sigma} w^{\sigma}
\alpha_0 +  (g^{\sigma} z^{\sigma})^{n-k-1}( g^{\sigma} w^{\sigma})^2 \, \alpha_1 ],
\ & (1\leqslant k < n) \\
e^{iN\tau\epsilon} \ [(g^{\sigma} z^{\sigma})^{N-n} g^{\sigma} w^{\sigma}\,
 \alpha_0     + g^{\sigma} z^{\sigma}(-\tau) \, \alpha_1 ] ,
\  &  (k = n) \\
         e^{iN\tau\epsilon} \ (g^{\sigma} z^{\sigma})^{N-k}
        g^{\sigma} w^{\sigma}   \alpha_0   \ & (n <k \leqslant N) .
\end{cases}
\]

Substitution of these expressions into (\ref{G-evol}) and (\ref{X-sigma}) allows to calculate
off-diagonal entries of the matrix $X^{\sigma}(N\tau)$ for $\zeta=\zeta^{(0, n)}$,
which correspond to the cross-terms involving $\alpha_0$ and $\alpha_1$.

Because of $|g^{\sigma} z^{\sigma}| < 1$ (Remark \ref{z-w-g-sigma}), these non-zero
off-diagonal entries will disappear when $N\to \infty$ for a fixed $n$.
Hence, in the long-time limit the composed subsystem $\mathcal{S} + \mathcal{S}_n$
evolves from the product of two initial  equilibrium states (\ref{st-S0-Sm}) to another product-state.
On the other hand, the cross-terms will not disappear in the limit
$N, n \to \infty$, when $N - n$ is fixed \cite{TZ}.
It is interesting that in this case the steady state of the subsystem $\mathcal{S}$ keeps a correlation with
subsystem $\mathcal{S}_n$ in the long-time limit.

\medskip

\noindent
\textit{Subsystem} $\mathcal{S}_{m} + \mathcal{S}_{n}$. We suppose that $1 \leqslant m < n \leqslant N $.
Then the initial state $\omega_{\mathcal{S}_{m} + \mathcal{S}_{n}}^0(\cdot)$ on
$\mathscr{A}(\mathscr{H}_{m}\otimes \mathscr{H}_{n}) \simeq \mathscr{A}(\mathscr{H}_{m})\otimes \mathscr{A}(\mathscr{H}_{n})$
of this composed subsystem is given by the partial trace
\begin{eqnarray}\label{st-Smn-Sm}
&&\omega_{\mathcal{S}_{m} + \mathcal{S}_{n}}^{0}(\widehat{w}(\alpha_{1})
\otimes\widehat{w}(\alpha_2))   =
\omega_{\rho}(\bigotimes_{k=0}^{m-1}\mathbb{1}\otimes
\widehat{w}(\alpha_{1})\otimes\bigotimes_{k=m+1}^{n-1}\mathbb{1}
\otimes \widehat{w}(\alpha_{2})\otimes \bigotimes_{k=n+1}^{N}\mathbb{1})
\nonumber \\
&& = \exp\Big[- \frac{|\alpha_{1}|^2}{4}\coth\frac{\beta}{2}\Big]
\exp\Big[- \frac{|\alpha_{2}|^2}{4} \coth\frac{\beta}{2}\Big]\ .
\end{eqnarray}
This is the characteristic function of the product-state corresponding to two isolated systems with the same temperature.

We define the vector $\zeta^{(m, n)} := {}^t(0, 0, \ldots , 0, \alpha_{1}, 0, \ldots , 0,
\alpha_{2}, 0, \ldots, 0) \in \C^{N + 1} $,
where $\alpha_{1}$ occupies the $(m+1)$th position and $\alpha_{2}$ occupies the $(n+1)$th position, then
\begin{equation}\label{st-Smn-t}
\omega_{\mathcal{S}_{m} + \mathcal{S}_{n}}^{N\tau}(\widehat{w}(\alpha_{1})\otimes\widehat{w}(\alpha_{2})) =
\omega_{\rho(N\tau)}(W(\zeta^{(m, n)})) \, .
\end{equation}
Again with help of Remark \ref{z-w-g-sigma}, we can calculate the components of
$ U_1^{\sigma}\ldots U_N^{\sigma} \ \zeta^{(m, n)}$ as
\begin{equation}\label{nmUk0}
    (U_1^{\sigma}\ldots U_N^{\sigma} \ \zeta^{(m, n)})_k =
\end{equation}
\[
     \begin{cases} \ e^{iN\tau\epsilon} \ (g^{\sigma} z^{\sigma} )^{m-1}\
    g^{\sigma} w^{\sigma}  [\alpha_{1}+(g^{\sigma} z^{\sigma})^{n-m} \alpha_2 ]
 \   \ & (k=0) \\
    e^{iN\tau\epsilon} \ (g^{\sigma} z^{\sigma} )^{m-k-1}
(g^{\sigma} w^{\sigma})^2 \,[ \alpha_{1} + (g^{\sigma} z^{\sigma} )^{n-m} \, \alpha_2 ]
\  \ & (1\leqslant k < m)
\\
e^{iN\tau\epsilon} \ [g^{\sigma} z^{\sigma}(-\tau) \, \alpha_{1}
      + (g^{\sigma} w^{\sigma})^2\, (g^{\sigma} z^{\sigma})^{n-m-1}\,
\alpha_2 ] \  \ & (k = m) \\
e^{iN\tau\epsilon} \   (g^{\sigma} z^{\sigma} )^{n-k-1}\,
       (g^{\sigma} w^{\sigma})^2\, \alpha_2 \   \ & (m<k<n)  \\
e^{iN\tau\epsilon} \ g^{\sigma} z^{\sigma}(-\tau) \, \alpha_{2} \  \ & (k = n) \\
0 \ \ & (n <k\leqslant N)
\end{cases}.
\]

The correlation between $\mathcal{S}_m$ and $\mathcal{S}_n$, i.e. the corresponding
off-diagonal elements of $X^{\sigma}(N\tau)$ are non-zero when $w\neq 0$,
and large for small $n-m$ and they decrease to zero as $n-m$ increase.
Note that in contrast to the case $\mathcal{S} + \mathcal{S}_{n}$ (\ref{0mUk0})
the last components $n< k \leqslant N$ in (\ref{nmUk0}) as well as the state (\ref{st-Smn-t})
do not depend on $N$.
This reflects the fact that correlation involving $\mathcal{S}_m$  and $\mathcal{S}_n$ via subsystem $\mathcal{S}$ is
switched off after the moment $t = n\tau$. If $w=0$, then (\ref{nmUk0}) implies that $X^{\sigma}(N\tau)$ is always diagonal
and that dynamics (\ref{st-Smn-t}) keeps $\mathcal{S}_{m} + \mathcal{S}_{n}$ uncorrelated.
\section{Time Evolution of Subsystems II}\label{RDM-subS}
The arguments of Section \ref{S-Sm} indicate that
the components in the subsystems $\mathcal{S} + \mathcal{S}_N + \ldots +
\mathcal{S}_{N-n}$ have large mutual correlations for small $n$ at $ t=N\tau$
even when $N$ large.
And those correlation seems asymptotically stable as $ N \to \infty$.

In this section, we consider the correlation among those components
simultaneously for product initial densities.
For this aim, let us divide the total system into two subsystems
$\mathcal{S}_{n, k}$ and $\mathcal{C}_{n, k}$ at the moment $t=k\tau$, where
\begin{equation}\label{S-nk}
\mathcal{S}_{n, k} = \mathcal{S} + \mathcal{S}_k  + \mathcal{S}_{k-1} +
\ldots + \mathcal{S}_{k-n+1}  ,
\end{equation}
and
\begin{equation}\label{C-nk}
       \mathcal{C}_{n, k} = \mathcal{S}_{N} + \ldots + \mathcal{S}_{k+1}
     \quad + \mathcal{S}_{k-n} + \ldots + \mathcal{S}_{1} .
\end{equation}
Here, $n \in \N$ is supposed to be fixed small and $N\in \N$ large enough.
We may imagine that the ``cavity" $\mathcal{S}$ and ``atoms" $\mathcal{S}_1, \ldots,
\mathcal{S}_N$ are lined as
\[
     \mathcal{S}_N, \ldots,  \mathcal{S}_{k+1}, \, \mathcal{S}, \, \mathcal{S}_k, \ldots,
  \mathcal{S}_{k-n+1}, \, \mathcal{S}_{k-n}, \ldots, \mathcal{S}_1 \,
\]
at this moment.
The interaction between $ \mathcal{S}$ and each of $\mathcal{S}_1 , \ldots, \mathcal{S}_k$
has already ended, and they are correlated.
While $ \mathcal{S}_{k+1}, \ldots,   \mathcal{S}_N$ have not interacted with $\mathcal{S}$, yet.
Let us regard that $\mathcal{S}_{n, k}$ is the ``state" at $t =k\tau$
of the time developing single object $\mathcal{S}_{\sim n}$.
That is,  $\mathcal{S}_{\sim n}$ has $\mathcal{S}, \mathcal{S}_k, \ldots,
\mathcal{S}_{k-n}$ as its components at the time $ t = k\tau$.
And it develops changing its components as well as the correlation among them.
As the time pass from $t=(k-1)\tau$ to $k\tau$, the ``atom" $\mathcal{S}_k$
enters into $\mathcal{S}_{\sim n}$ and the ``atom" $\mathcal{S}_{k-n}$
leaves from $\mathcal{S}_{\sim n}$.
It is also possible to regard $\mathcal{S}_{\sim n}$ as the view from the window
which is made to look the ``cavity" and the $n$ ``atoms" just have interacted with
the ``cavity".

We are interested in $\mathcal{S}_{\sim n}$, since it might be interpreted
as a simplified mathematical model of physical objects in equilibrium with the reservoir
or of metabolizing life forms which maintain their life by interacting with
the environment, i.e., the macroscopic many body systems which are
macroscopically stable but exchange their constituent particles as well as
energy with the reservoir microscopically.

Below we consider the large-time asymptotic behavior of state for
$\mathcal{S}_{\sim n}$,
i.e., for the subsystem $\mathcal{S}_{n, k}$ with fixed $n$ and large
and variable $k$ for the initial state (\ref{state-S-C1}) with general
density matrices $\rho_0, \rho_1 \in \mathfrak{C}_1(\mathscr{F})$.

To express the state of $\mathcal{S}_{\sim n}$ at $t = k\tau$,
we decompose the Hilbert space $\mathscr{H}$ into a tensor product of two
Hilbert spaces
\[
            \mathscr{H} = \mathscr{H}_{\mathcal{S}_{n, k}} \bigotimes
        \mathscr{H}_{\mathcal{C}_{n, k}} \, .
\]
Here $\mathscr{H}_{\mathcal{S}_{n, k}}$ is the Hilbert space for the
subsystem (\ref{S-nk}) and $\mathscr{H}_{\mathcal{C}_{n, k}}$ for (\ref{C-nk}):
\begin{equation}\label{H-s-H-c}
          \mathscr{H}_{\mathcal{S}_{n, k}} = \mathscr{H}_0 \bigotimes
          \Big( \bigotimes_{j= k-n+1}^k\mathscr{H}_{j} \Big) , \qquad
           \mathscr{H}_{\mathcal{C}_{n, k}} = \Big( \bigotimes_{j=1}^{k-n}
           \mathscr{H}_j\Big) \bigotimes
         \Big( \bigotimes_{l= k+1}^N\mathscr{H}_{l} \Big) .
\end{equation}
If $\rho \in\mathfrak{C}_1(\mathscr{H})$ is the initial density matrix of the total system
$\mathcal{S}_{n, k}+ \mathcal{C}_{n, k}$,
the reduced density matrix $\rho_{\mathcal{S}_{\sim n}}(k\tau)$ of
$\mathcal{S}_{\sim n}$ at $t=k\tau $  is given by the partial trace
\begin{equation}\label{S-rho-k}
     \rho_{\mathcal{S}_{\sim n}}(k\tau) = \Tr_{\mathscr{H}_{\mathcal{C}_{n, k}}}
     (T^{\sigma}_{k\tau,0} \, \rho)
     = \Tr_{\mathscr{H}_{c_1}} \big( \Tr_{\mathscr{H}_{c_2}}
         (T^{\sigma}_{k\tau,0} \, \rho) \big) \ ,
\end{equation}
for $k \geqslant n$ as in (\ref{P-Tr}), where we decompose
$\mathscr{H}_{\mathcal{C}_{n, k}}$ as
\[
          \mathscr{H}_{\mathcal{C}_{n, k}} = \mathscr{H}_{c_1} \bigotimes \mathscr{H}_{c_2} \,,  \quad
           \mathscr{H}_{c_1} = \bigotimes_{j=1}^{k-n}\mathscr{H}_j \,, \quad
          \mathscr{H}_{c_2} = \bigotimes_{l= k+1}^N\mathscr{H}_{l} \, .
\]
\subsection{Preliminaries}
\label{preliminaries-bis}
Here we introduce notations and definitions to study evolution of subsystems in somewhat more general setting than
in the previous sections.

In order to avoid the confusion caused by the fact that every $\mathscr{H}_j$ coincides
with $\mathscr{F}$ in our case,
we treat the Weyl algebra on the subsystem and the corresponding reduced density matrix of
$ \rho \in \mathfrak{C}_{1}(\mathscr{H}) $ in the following way. On the Fock space $ \mathcal{F}^{\otimes(m+1)} $ for
$m = 0,1, \ldots , N$, we define the Weyl operators
\begin{equation}\label{W-m-tilde}
           W_m(\zeta) := \exp \Big(i \ \frac{\langle \zeta, \tilde b \rangle_{m+1}
      + \langle \tilde b, \zeta \rangle_{m+1}}{\sqrt 2} \Big) \  ,
\end{equation}
where $\zeta \in \C^{m+1}$, $ \tilde b_0, \ldots, \tilde b_m$ and
$\tilde b^*_0, \ldots, \tilde b_m^*$ are the annihilation and the creation
operators in $ \mathcal{F}^{\otimes(m+1)} $,
which are constructed as in (\ref{k-bosons}) satisfying the corresponding CCR and
\[
     \langle \zeta, \tilde b \rangle_{m+1} = \sum_{j=0}^m \bar{\zeta_j}
 \tilde b_j, \qquad  \langle \tilde b, \zeta \rangle_{m+1}
     = \sum_{j=0}^m \zeta_j\tilde{b}^*_j .
\]
By $\mathscr{A}(\mathscr{F}^{\otimes(m+1)})$, we denote the $C^*$-algebra
generated by the Weyl operators (\ref{W-m-tilde}).

To discuss the dynamics of our open system, it is convenient to introduce the \textit{modified} Weyl operators
(cf. Proposition \ref{one-step-dualT})
\begin{equation}\label{m-W-m}
           W_m^{\sigma}(\zeta) := \exp \Big[\frac{\sigma_-+\sigma_+}{4(\sigma_- - \sigma_+)}
            \langle \zeta, \zeta \rangle_{m+1}\Big] \, W_m(\zeta) \ ,
\end{equation}
for $ m = 0 , 1, 2, \ldots $, where $ \zeta \in \C^{m+1}$ and $\langle \, \cdot \, , \, \cdot \, \rangle_{n+1}$ denotes
the inner product on $\C^{m+1}$. We also use the notation
\begin{equation}\label{m-w}
         \widehat w^{\sigma}(\theta) := W_0^{\sigma}(\theta)
         \qquad \mbox{for} \quad \theta \in \C .
\end{equation}
Below, we adopt the abreviations:
\begin{equation}\label{AnCn}
         \mathscr{A}^{(m)} = \mathscr{A}(\mathscr{F}^{\otimes(m+1)}) \quad \mbox{and} \quad
          \mathscr{C}^{(m)} =  \mathfrak{C}_1(\mathscr{F}^{\otimes(m+1)})
\end{equation}
for the Weyl $C^*$ algebra on $\mathscr{F}^{\otimes(m+1)}$  and the algebra of all trace class operators on
$\mathscr{F}^{\otimes(m+1)}$ for $ m = 0, 1, 2, \ldots $, respectively.
Note that the bilinear form
\begin{equation}\label{dualCA}
      \langle \ \cdot \ | \ \cdot  \ \rangle_m : \mathscr{C}^{(m)} \times \mathscr{A}^{(m)}
     \ni (\rho, A) \mapsto \Tr[\rho A] \in \C
\end{equation}
yields the dual pair $(\mathscr{C}^{(m)}, \mathscr{A}^{(m)} )$.
Indeed, the following properties hold:

\medskip

 (i) \; \quad  $\langle \rho \, |\,  A \rangle_m = 0 \mbox{ for every } A \in \mathscr{A}^{(m)}$
 \quad implies \quad $ \rho = 0 $;

\smallskip

(ii) \, \quad $\langle \rho \, |\, A \rangle_m = 0 $ for every  $ \rho \in \mathscr{C}^{(m)}$
 \quad implies \quad $ A =0 $;

\smallskip

(iii)  \quad $ | \langle \rho \, |\, A \rangle_m | \leqslant \|\rho \|_{\mathfrak{C}_1}
      \| A \|_{\mathcal{L}} $.

\medskip

\noindent These properties are a direct consequence of the fact that $\mathscr{A}^{(m)}$ is weakly dense in
$\mathcal{L}(\mathscr{F}^{\otimes(m+1)})$ the dual space of $\mathscr{C}^{(m)}$.
Below we shall use  the topology $\sigma(\mathscr{C}^{(m)}, \mathscr{A}^{(m)})$
induced by the dual pair $(\mathscr{C}^{(m)}, \mathscr{A}^{(m)})$ on $\mathscr{C}^{(m)}$.
We refer to it as the weak${}^*$-$\mathscr{A}^{(m)}$ topology,
see e.g. \cite{Ro}, \cite{BR1}.

\medskip

Note that for the initial normal \textit{product} state (\ref{state-S-C1}) the calculation of the partial trace over
$\mathscr{H}_{c_2}$ in (\ref{S-rho-k}) is straightforward:
\begin{equation}\label{ptrace2}
       \Tr_{\mathscr{H}_{c_2}}(T^{\sigma \, (N)}_{k\tau,0} \, \bigotimes_{j=0}^N\rho_j)
       = T^{\sigma \, (k)}_{k\tau,0} \, \bigotimes_{j=0}^k\rho_j \, .
\end{equation}
Here $ T^{\sigma \, (m)}_{k\tau,0}$ stands for the evolution map (\ref{T-sigma-t})
on $\mathscr{C}^{(m)}$, for $ k \leqslant m \leqslant N $.

\medskip

To check (\ref{ptrace2}), it is enough to show
\begin{equation}\label{ptN-k}
    \langle T^{\sigma \, (N)}_{k\tau,0} \, \bigotimes_{j=0}^N \rho_j | W_k(\zeta) \otimes \mathbb{1} \rangle_N =
      \langle T^{\sigma \, (k)}_{k\tau,0} \, \bigotimes_{j=0}^k \rho_j | W_k(\zeta) \rangle_k
\end{equation}
for any $\zeta \in \C^{k+1}$, where $\mathbb{1} $ is the unit in algebra $\mathscr{A}^{(N-k-1)}$.
Let $\tilde\zeta \in \C^{N+1}$ be defined by $ \tilde\zeta_j = \zeta_j$ for
$ 0\leqslant j \leqslant k$, $\tilde\zeta_j =0$ for $ k< j \leqslant N$.
Then $ W_k(\zeta) \otimes \mathbb{1} = W_N(\tilde\zeta)$ holds.
Remark \ref{z-w-g-sigma} readily yields
\[
     U_1^{\sigma(N)} \ldots U_k^{\sigma(N)} \, \tilde\zeta
   = (U_1^{\sigma(k)} \ldots U_k^{\sigma(k)} \, \zeta)\widetilde{\quad } \, .
\]
Together with {(\ref{T*-k-W}), it follows that
\[
          T^{\sigma \, (N) *}_{k\tau,0} (\,W_N(\tilde\zeta) \, ) =
           \Big( T^{\sigma \, (k) *}_{k\tau,0} \, W_k(\zeta) \Big) \otimes \mathbb{1} ,
\]
which implies
\begin{equation}\label{ptN-k*}
     \langle  \bigotimes_{j=0}^N \rho_j \, | \, T^{\sigma \, (N) *}_{k\tau,0} (\,W_k(\zeta)
       \otimes \mathbb{1} )\rangle_N =
      \langle\bigotimes_{j=0}^k \rho_j \, | \,  T^{\sigma \, (k) *}_{k\tau,0} \, W_k(\zeta)
      \rangle_k \, .
\end{equation}
This proves (\ref{ptN-k}) and thereby the assertion (\ref{ptrace2}).

Here we have used the notation $ U_{\ell}^{\sigma(k)}$ for the $(k+1)\times (k+1)$
matrix whose components are given by
\begin{equation}\label{U-sigma-m}
        (U_{\ell}^{\sigma(k)})_{ij} = \begin{cases}
      e^{i\tau\epsilon}g^{\sigma}(\tau) (\delta_{j0}z^{\sigma}(\tau)
   + \delta_{j{\ell}}w^{\sigma}(\tau) )
       &  ( i = 0 ) \\
    e^{i\tau\epsilon}g^{\sigma}(\tau) (\delta_{j0}w^{\sigma}(\tau)
    + \delta_{j{\ell}}z^{\sigma}(-\tau) )
       &  ( i = {\ell} ) \\
       e^{i\tau\epsilon}\delta_{ij}   &  ( \mbox{otherwise} )
      \end{cases},
\end{equation}
for $ {\ell} = 1, 2, \ldots, k$ (c.f. Remark \ref{z-w-g-sigma} ).
Then the one step evolution $ T_{\ell}^{\sigma(k)}$ on $\mathscr{C}^{(k)}$ is given by
\[
      \langle T_{\ell}^{\sigma(k)} \rho \, |\,  W_k(\zeta)  \rangle_k =
          \langle \rho \, |\,  T_{\ell}^{\sigma(k)*} W_k(\zeta)  \rangle_k  \,
\]
where
\begin{equation}\label{T*Wm}
          T_{\ell}^{\sigma(k)*} W_k(\zeta)  = \exp \Big[- \frac{\sigma_-+\sigma_+}{4(\sigma_- - \sigma_+)}
        \big(  \langle \zeta, \zeta \rangle_{k+1} - \langle U_{\ell}^{\sigma(k)} \zeta,
          U_{\ell}^{\sigma(m)}  \zeta \rangle_{k+1} \big)  \Big]
    W_k(U_{\ell}^{\sigma(k)}  \zeta) \,,
\end{equation}
$\rho \in \mathscr{C}^{(k)}$ and $ \zeta \in \C^{k+1}$
(see Proposition \ref{one-step-dualT}).

\medskip

To calculate the partial trace (\ref{S-rho-k}) with respect to $\mathscr{H}_{c_1}$, we introduce the imbedding:
\begin{equation}\label{r}
         r_{m+1, m}: \C^{m+1} \ni \zeta = \begin{pmatrix} \zeta_0 \\ \zeta_1 \\ \zeta_2 \\
            \cdot \\  \cdot \\  \cdot \\  \zeta_m \end{pmatrix} \longmapsto
              \begin{pmatrix} \zeta_0 \\ 0 \\ \zeta_1 \\ \zeta_2 \\
            \cdot \\  \cdot \\  \cdot \\  \zeta_m \end{pmatrix} = r_{m+1,m}\zeta \in  \C^{m+2}
\end{equation}
for $ m = 0, 1, 2, \ldots, N$ and the {partial trace} over the \textit{second} component
$ R_{m, m+1} : \mathscr{C}^{(m+1)} \rightarrow \mathscr{C}^{(m)} $  characterised by
\begin{equation}\label{R}
       \langle  R_{m,m+1}\rho | \widehat w(\zeta_0) \otimes \widehat w(\zeta_1) \otimes \ldots \otimes
      \widehat w(\zeta_m)  \rangle_m = \langle \rho | \widehat w(\zeta_0) \otimes \mathbb{1}
     \otimes \widehat w(\zeta_1) \otimes \ldots \otimes  \widehat w(\zeta_m)  \rangle_{m+1}
\end{equation}
for $\rho \in \mathscr{C}^{(m+1)}$.
Therefore, its dual operator $R_{m,m+1}^{*} $ has the expression:
\begin{equation}\label{R*Wr}
R_{m,m+1}^*W_m(\zeta) = W_{m+1}(r_{m+1,m}\zeta)  \quad
       \mbox{for} \quad \zeta \in \C^{m+1} \, .
\end{equation}
\begin{lem}\label{Ur}
For $ m \in \N$ and $ {\ell} = 1, 2, \ldots, m$,
\begin{equation}\label{Ur-rU}
      U_{{\ell}+1}^{\sigma(m+1)}r_{m+1, m} = r_{m+1,m}U_{\ell}^{\sigma(m)}  \ ,
\end{equation}
holds.
\end{lem}
{\sl Proof  } : In fact, for the vector $\displaystyle \zeta = {}^t
( \zeta_0,  \zeta_1, \cdots, \zeta_m) \in \C^{m+1}$, one obtains
\[
     ( U_{{\ell}+1}^{\sigma(m+1)}r_{m+1,m}\zeta)_j  = (r_{m+1,m}U_{\ell}^{\sigma(m)}\zeta)_j
\]
\[
    = \begin{cases} e^{i\tau\epsilon}g^{\sigma}(\tau)(z^{\sigma}(\tau)\zeta_0
          + w^{\sigma}(\tau)\zeta_{\ell} ) & ( j=0 )  \\
                 0 & ( j=1 )  \\
       e^{i\tau\epsilon}\zeta_{j-1} & (2 \leqslant j \leqslant {\ell}) \\
       e^{i\tau\epsilon}g^{\sigma}(\tau)(w^{\sigma}(\tau)\zeta_0
          + z^{\sigma}(-\tau)\zeta_{\ell} ) & ( j = {\ell}+1 )  \\
      e^{i\tau\epsilon}\zeta_{j-1} & ( {\ell} + 2 \leqslant j \leqslant m+1 ) \end{cases}
\]
by explicit calculations.
This proves the claim (\ref{Ur-rU}). \hfill $\square$

\medskip

For $ k \in \N$ and $ m= 0, 1, 2, \ldots, k-1$, let the maps $r_{k,m}: \C^{m+1} \to \C^{k+1}$ and
$R_{m,k} : \mathscr{C}^{(k)} \to \mathscr{C}^{(m)}$ be defined by composition of the
one-step maps (\ref{r}), (\ref{R}):
\[
       r_{k,m} = r_{k,k-1}\circ r_{k-1,k-2}\circ \ldots \circ r_{m+1,m} \ ,
\]
and
\[
       R_{m,k} = R_{m, m+1}\circ R_{m+1,m+2}\circ \ldots \circ R_{k-1,k}  \,,
\]
respectively.
This definition together with (\ref{R}) and (\ref{R*Wr}) imply that
$R^{\ast}_{m,k}: \mathscr{A}^{(m)} \rightarrow \mathscr{A}^{(k)}$ and
\begin{equation}\label{m-k-map}
R^{\ast}_{m,k} \, \widehat w(\zeta_0) \otimes \widehat w(\zeta_1) \otimes \ldots \otimes \widehat w(\zeta_m) =
\widehat w(\zeta_0) \otimes \mathbb{1} \otimes \ldots \otimes \mathbb{1} \otimes\widehat w(\zeta_1) \otimes \ldots
\otimes \widehat w(\zeta_m) \ .
\end{equation}

Hence, by (\ref{R}) the map $R_{m,k}$, which is predual to (\ref{m-k-map}), acts as the partial trace
over the components with indices $j=1, 2, \ldots, k-m$
of the tensor product $\bigotimes_{j=0}^k \rho_j \in  \mathscr{C}^{(k)}$.
Therefore, the map $R_{n,k}$ coincides with the partial trace $\Tr_{c_1}$
in (\ref{S-rho-k}).
Then $R_{n,k}$ combined with (\ref{ptrace2}) gives the expression
\begin{equation}\label{ptrace3}
\rho_{\mathcal{S}_{\sim n}}(k\tau) = R_{n,k }T^{\sigma(k)}_{k\tau, 0}
     (\bigotimes_{j=0}^k\rho_j) \quad \mbox{for}
 \quad    k \geqslant n+1 \, .
\end{equation}

We summarise the action of the above maps (\ref{r}), (\ref{R*Wr})-(\ref{T*Wm}) on the modified Weyl operators (\ref{m-W-m}) by
\begin{lem} \label{RTW}
Let $k \in \N$. Then,
\begin{eqnarray}
{\rm (i)}&\hspace{3cm}  R_{m,m+1}^*(W_m^{\sigma}(\zeta)) =
          W_{m+1}^{\sigma}(r_{m+1,m}\zeta) \, , \hspace{3cm} \phantom{L}\\
{\rm (ii)}& R_{m,m+k}^*(W_m^{\sigma}(\zeta)) =
          W_{m+k}^{\sigma}(r_{m+k,m}\zeta)
\end{eqnarray}
holds for $m =0, 1, 2, \ldots$, $\zeta \in \C^{m+1}$; and
\begin{eqnarray}
{\rm (iii)}& T_{\ell}^{\sigma(m)*}(W_m^{\sigma}(\zeta))
        = W_m^{\sigma}(U_{\ell}^{\sigma(m)}\zeta) \,, \\
{\rm (iv)}&\hspace{8mm}  T_1^{\sigma(m)*}T_2^{\sigma(m)*}\ldots
    T_{\ell}^{\sigma(m)*}(W_m^{\sigma}(\zeta))
        = W_m^{\sigma}(U_1^{\sigma(m)}\ldots
     U_{{\ell}-1}^{\sigma(m)}U_{\ell}^{\sigma(m)}\zeta) \, , \hspace{7mm}\phantom{L} \\
{\rm (v)}&  U_{{\ell}+k}^{\sigma(m+k)}r_{m+k,m} = r_{m+k,m}
              U_{\ell}^{\sigma(m)}
\end{eqnarray}
holds for $m \in \N$, $\zeta \in \C^{m+1}$ and $ {\ell} = 1, 2, \ldots, m $.
\end{lem}
Note that the claim (v) in the above is an obvious extension of Lemma \ref{Ur}.
This lemma yields the following statement.
\begin{lem}\label{RT}
For $m, k \in \N $, $ {\ell} = 1, 2, \ldots, m$,
\begin{equation}\label{R*T*}
      R^*_{m, m+k}T_{\ell}^{\sigma(m)*} = T_{{\ell}+k}^{\sigma(m+k)*}
      R^*_{m,m+k}
\end{equation}
holds on $ \mathscr{A}^{(m)}$. Therefore
\begin{equation}\label{R*T*1}
      R_{m, m+k}T_{{\ell}+k}^{\sigma(m+k)} = T_{\ell}^{\sigma(m)}
      R_{m,m+k}
\end{equation}
and
\[
       R_{m, m+k}T_{k}^{\sigma(m+k)} T_{k-1}^{\sigma(m+k)}
      \ldots T_{1}^{\sigma(m+k)}
\]
\begin{equation}\label{R*T*2}
      = (R_{m,m+1}T_{1}^{\sigma(m+1)})\ldots
      (R_{m+k-2,m+k-1}T_{1}^{\sigma(m+k-1)})
     (R_{m+k-1,m+k}T_{1}^{\sigma(m+k)})
\end{equation}
hold on $\mathscr{C}^{(m+k)}$.
\end{lem}
{\sl Proof : } The identity (\ref{R*T*}) follows from Lemma \ref{RTW} by considering the
action on the modified Weyl operators.
By taking its adjoint, (\ref{R*T*1}) follows.
A simple application of induction over $k$ yields the last
identity.  \hfill $\square$

\medskip

Let us concentrate on the evolution of the subsystem $\mathcal{S}$, first.
To this aim, we introduce the map $ \; \mathcal{T}[ \, \cdot \, | \, \cdot \, ] :
\mathscr{C}^{(0)} \times \mathscr{C}^{(0)}  \to \mathscr{C}^{(0)} \; $
to express the {\it one-step} evolution
\begin{equation}\label{T[|]}
     \mathcal{T}[ \rho_0 | \rho_1 ] = R_{0,1}T_1^{\sigma(1)}
   (\rho_0 \otimes \rho_1) \quad \mbox{for} \quad
    \rho_0, \rho_1 \in \mathscr{C}^{(0)} \ ,
\end{equation}
of the density matrix $\rho_0$ under the influence of $\rho_1$, see (\ref{ptrace3}).
We also denote by
\begin{equation}\label{T[-]}
    \mathcal{T}[ \rho ] := e^{-i\epsilon\tau a^*a}\rho
     e^{i\epsilon\tau a^*a} \quad \mbox{for} \quad
    \rho \in \mathscr{C}^{(0)} \ ,
\end{equation}
the ``\textit{free}" one-step evolution of density matrix corresponding to any of
subsystems $\mathcal{S}_{k}$, c.f. (\ref{Ham-n}).
Then one obtains the following assertion.
\begin{lem}\label{T[]} For any $ {\ell}, m \in \N$ fulfilling ${\ell}\leqslant m$,
$\zeta \in \C^{m}$,  $\theta \in \C$
and $\rho_0 , \rho_1, \ldots, \rho_{\ell} \in \mathscr{C}^{(0)}$,
the following properties hold:
\begin{eqnarray}
{\rm (i)}&  \mathcal{T}^{*\otimes m}[W_{m-1}^{\sigma}(\zeta)] = W_{m-1}^{\sigma}
      (e^{i\epsilon\tau}\zeta) ,
       (\mathcal{T}^{-1})^{*\otimes m}[W_{m-1}^{\sigma}(\zeta)] = W_{m-1}^{\sigma}
      (e^{-i\epsilon\tau}\zeta) \;  ;
 \notag \\
{\rm (ii)}&  \mathcal{T}R_{0,1} = R_{0,1}
      \mathcal{T}^{\otimes 2} ; \notag\\
{\rm (iii)}&     \langle  \mathcal{T}[ \rho_0 | \rho_1 ]  \, |\,
         \widehat w^{\sigma}(\theta) \rangle_0 =
         \langle \rho_0 \, |\, \widehat w^{\sigma}(e^{i\epsilon\tau}
   g^{\sigma}(\tau)z^{\sigma}(\tau) \theta)  \rangle_0
         \langle  \rho_1 \, |\, \widehat w^{\sigma}(e^{i\epsilon\tau}
   g^{\sigma}(\tau)w^{\sigma}(\tau) \theta)  \rangle_0 \; ; \notag \\
{\rm (iv)}& \mathcal{T}^{\otimes (m+1)}  T_{\ell}^{\sigma(m)}
= T_{\ell}^{\sigma(m)}\mathcal{T}^{\otimes (m+1)}, \quad
      (\mathcal{T}^{-1})^{\otimes (m+1)}  T_{\ell}^{\sigma(m)}
= T_{\ell}^{\sigma(m)}(\mathcal{T}^{-1})^{\otimes (m+1)}  \;;  \notag  \\
{\rm (v)}& \mathcal{T}\big(\mathcal{T}[\rho_0 | \rho_1 ] \big) =
        \mathcal{T}[\mathcal{T}\rho_0 | \mathcal{T}\rho_1 ],
        \quad
        \mathcal{T}^{-1}\big( \mathcal{T}[\rho_0 | \rho_1 ] \big) =
            \mathcal{T}[\mathcal{T}^{-1}\rho_0 |
        \mathcal{T}^{-1}\rho_1 ] \; ; \notag \\
{\rm (vi)}&  R_{{\ell}-1,{\ell}}T_1^{\sigma({\ell})}[\rho_0 \otimes \rho_1 \otimes \ldots \otimes \rho_{\ell}] =
\mathcal{T}[\rho_0 | \rho_1 ] \otimes
          \mathcal{T}[\rho_2] \otimes \ldots \otimes
       \mathcal{T}[\rho_{\ell}]. \notag
\end{eqnarray}
Here $(\cdot)^{\otimes(m+1)}$ denotes the $(m+1)$-fold tensor product of the corresponding
operator $(\cdot)$.
\end{lem}
{\sl Proof : } (i) Since
\[
    \langle \rho  \, |\, \mathcal{T}^*
   [\widehat w^{\sigma}(\theta)]\rangle_0 =
   \langle \mathcal{T}\rho  \, |\, [\widehat w^{\sigma}(\theta)]\rangle_0
\]
\[
    = \Tr (\rho e^{i\epsilon\tau a^*a}\widehat w^{\sigma}(\theta)
   e^{-i\epsilon\tau a^*a}) = \Tr (\rho \widehat w^{\sigma}(
     e^{i\epsilon\tau }\theta)) = \langle \rho  \, |\,
   \widehat w^{\sigma}(e^{i\epsilon\tau }\theta) \rangle_0
\]
holds for $ \rho \in  \mathscr{C}^{(0)}$, we obtain the desired equality for $m=1$.
The equalities for $m > 1$ follow from (\ref{Wz}) and the definition of tensor product
of $(\mathcal{T}^{\pm 1})^*$.

\noindent(ii) Taking into account (i) of Lemma \ref{RTW} and the above (i), we get
\[
  (\mathcal{T}R_{0,1})^* \widehat w^{\sigma}(\theta)
   = R_{0,1}^* \mathcal{T}^* \widehat w^{\sigma}(\theta)
   = R_{0,1}^* \widehat w^{\sigma}(e^{i\epsilon\tau}\theta)
   = W_1^{\sigma}(e^{i\epsilon\tau}r_{1,0}\theta)
\]
\[
      =   \mathcal{T}^{\otimes2 *} W_1^{\sigma}(r_{1,0}\theta)
      = \mathcal{T}^{\otimes2 *}R_{0,1}^*  \widehat w^{\sigma}(\theta)
      =  (R_{0,1}\mathcal{T}^{\otimes2})^* \widehat w^{\sigma}(\theta).
\]
(iii) By virtue of (\ref{r}) and (\ref{U-sigma-m}) one obtains
\[
     U_1 ^{\sigma(1)}r_{1,0}\theta = \begin{pmatrix}
   e^{i\tau\epsilon}g^{\sigma}z^{\sigma},
      e^{i\tau\epsilon}g^{\sigma}w^{\sigma} \\
     e^{i\tau\epsilon}g^{\sigma}w^{\sigma},
     e^{i\tau\epsilon}g^{\sigma} z^{\sigma}(-\tau)
      \end{pmatrix}
       \begin{pmatrix} \theta \\ 0 \end{pmatrix}
\]
\[
       = \begin{pmatrix}
   e^{i\tau\epsilon}g^{\sigma}z^{\sigma}\theta \\
     e^{i\tau\epsilon}g^{\sigma}w^{\sigma} \theta    \end{pmatrix} ,
\]
which impllies
\[
     \langle  \mathcal{T}[ \rho_0 | \rho_1 ]  \, |\,
         \widehat w^{\sigma}(\theta) \rangle_0 =
         \langle  R_{0,1}T_1^{\sigma(1)}[ \rho_0 \otimes
    \rho_1 ] \, |\, \widehat w^{\sigma}(\theta) \rangle_0
\]
\[
     = \langle \rho_0 \otimes \rho_1 \, |\, T_1^{\sigma(1)*}
         R_{0,1}^* \widehat w^{\sigma}(\theta)\rangle_1
      = \langle \rho_0 \otimes \rho_1 \, |\, W_1^{\sigma}(
     U_1^{\sigma(1)}r_{1,0}\theta)\rangle_1
\]
\begin{equation}\label{one-step-id}
         = \langle \rho_0 \, |\, \widehat w^{\sigma}(e^{i\epsilon\tau}
   g^{\sigma}z^{\sigma} \theta)  \rangle_0
         \langle  \rho_1 \, |\, \widehat w^{\sigma}(e^{i\epsilon\tau}
   g^{\sigma}w^{\sigma} \theta)  \rangle_0 \, .
\end{equation}
(iv) By applying the adjoint operators
\[
      (\mathcal{T}^{\pm 1})^{\otimes (m+1)* } , \quad
        T_{\ell}^{\sigma(m)*}
\]
to the modified Weyl operators, we get
\[
   (\mathcal{T}^{\pm1})^{\otimes (m+1)*}  T_{\ell}^{\sigma(m)*}
  = T_{\ell}^{\sigma(m)*}(\mathcal{T}^{\pm 1})^{\otimes (m+1)*}
\]
from (iii) of Lemma \ref{RTW} and the above (i).
Then, duality derives the assertion.

\noindent (v) These identities follow from the above (ii), (iv) and the definition (\ref{T[|]}).

\noindent (vi) Let $\zeta \in \C^{\ell}$.
By (\ref{r}) and (\ref{U-sigma-m}), we obtain
\[
     U_1 ^{\sigma({\ell})}r_{{\ell},{\ell}-1}\zeta
       = e^{i\tau\epsilon \ \ t}
   ( g^{\sigma}z^{\sigma}\zeta_0, \,  g^{\sigma}w^{\sigma} \zeta_0,   \,
          \zeta_1, \, \cdots,  \, \zeta_{{\ell}-1}) \, ,
\]
where ${}^t ( \cdots )$ is the vector transposition.
Then we get
\[
    \langle  R_{{\ell}-1,{\ell}}T_1^{\sigma({\ell})} ( \rho_0 \otimes
    \rho_1 \otimes \ldots \otimes \rho_{\ell})  \, |\,
      W_{{\ell}-1}^{\sigma}(\zeta) \rangle_{\ell -1}
\]
\[
     = \langle \rho_0 \otimes \rho_1 \otimes \ldots
    \otimes \rho_{\ell} \, |\, T_1^{\sigma({\ell})*}
         R_{{\ell}-1,{\ell}}^* W_{{\ell}-1}^{\sigma}(\zeta)\rangle_{\ell}
      = \langle \rho_0 \otimes \rho_1\otimes \ldots
      \otimes \rho_{\ell} \, |\,  W_{\ell}^{\sigma}(
     U_1^{\sigma({\ell})}r_{{\ell}, {\ell}-1}\zeta)\rangle_{\ell}
\]
\[
       = \langle \rho_0 \, |\, \widehat w^{\sigma}(e^{i\epsilon\tau}
   g^{\sigma}(\tau)z^{\sigma}(\tau) \zeta_0)  \rangle_0
         \langle  \rho_1 \, |\, \widehat w^{\sigma}(e^{i\epsilon\tau}
   g^{\sigma}(\tau)w^{\sigma}(\tau) \zeta_0)  \rangle_0
 \langle \rho_2 \, |\, \widehat w^{\sigma}(e^{i\epsilon\tau}
   \zeta_1)  \rangle_0 \ldots \langle \rho_{\ell} \, |\,
   \widehat w^{\sigma}(e^{i\epsilon\tau} \zeta_{{\ell}-1})  \rangle_0
\]
\[
    = \langle \mathcal{T}[\rho_0 | \rho_1 ] \, |\,
       \widehat w^{\sigma}(\zeta_0)  \rangle_0
      \langle \rho_2  \, |\, \mathcal{T}^*[w^{\sigma}(\zeta_1)]\rangle_0
      \ldots  \langle
      \rho_{\ell} \, |\, \mathcal{T}^*[w^{\sigma}(\zeta_{{\ell}-1})]\rangle_0
\]
\[
   = \langle \mathcal{T}[\rho_0 | \rho_1 ] \otimes
          \mathcal{T}[\rho_2] \otimes \ldots \otimes
       \mathcal{T}[\rho_{\ell}]  \, |\, W_{{\ell}-1}^{\sigma}(\zeta)\rangle_{\ell -1} \,,
\]
where we used (iii) and (i) at the fourth equality.
These finish the proof of the lemma.  \hfill $\square$

Note that (\ref{one-step-id}) coincides with (\ref{one-step-evol1}).

Next, we consider the multi-step evolution for the
subsystem $\mathcal{S}$.
To this aim, we define $ \mathcal{T}^{(k)} : \mathscr{C}^{(0) (k+1)}
\to \mathscr{C}^{(0)} $ by
\begin{equation}\label{defTk}
       \mathcal{T}^{(k)}[ \rho_0 | \rho_1, \ldots, \rho_k ]
       = R_{0,k}T_k^{\sigma(k)}  T_{k-1}^{\sigma(k)}
  \ldots T_1^{\sigma(k)}(\rho_0 \otimes \rho_1 \otimes
    \ldots \otimes \rho_k) ,
\end{equation}
for $ k \in \N $ and
$ \rho_0, \rho_1, \ldots, \rho_k \in \mathscr{C}^{(0)}$,
c.f. (\ref{ptrace3}), (\ref{T[|]}).

The following lemma holds.
\begin{lem}\label{Tk[]} For any $ \theta \in \C $, $k \in \N$ and
$\rho_0 , \rho_1, \ldots, \rho_{k}, \ldots \in \mathscr{C}^{(0)}$,
the following properties hold:
\begin{eqnarray*}
{\rm (i)}&&  \mathcal{T}^{(1)}[ \rho_0 | \rho_1 ] = \mathcal{T}
    [ \rho_0 | \rho_1 ]\qquad ; \\
{\rm (ii)}&&  \mathcal{T}^{(k+1)}[ \rho_0 | \rho_1, \ldots,
  \rho_{k+1}] = \mathcal{T}[ \mathcal{T}^{(k)}[ \rho_0 |
   \rho_1, \ldots, \rho_k] | \mathcal{T}^k\rho_{k+1}]  \\
  &&  =   \mathcal{T}^{(k)}[ \mathcal{T}[ \rho_0 | \rho_1] |
  \mathcal{T}\rho_2, \ldots, \mathcal{T}\rho_{k+1}] \quad ; \\
{\rm (iii)}&& \mathcal{T}\Big(\mathcal{T}^{(k)}[\rho_0 |
   \rho_1 , \ldots, \rho_k ] \Big) =
        \mathcal{T}^{(k)}[\mathcal{T}\rho_0 |
   \mathcal{T}\rho_1 , \ldots, \mathcal{T}\rho_k ] \quad ; \\
  &&     \mathcal{T}^{-1}\Big(\mathcal{T}^{(k)}[\rho_0 | \rho_1 , \ldots, \rho_k ] \Big)
    = \mathcal{T}^{(k)}[\mathcal{T}^{-1}\rho_0 |
   \mathcal{T}^{-1}\rho_1 , \ldots, \mathcal{T}^{-1}\rho_k ]  \quad ; \\
{\rm (iv)}&&  R_{m,k+m}T_k^{\sigma(k+m)} \ldots T_1^{\sigma(k+m)}
[\rho_0 \otimes \rho_1 \otimes \ldots \otimes \rho_{k+m}] \\
  && = \mathcal{T}^{(k)}[\rho_0 | \rho_1, \ldots, \rho_k]
   \otimes \mathcal{T}^k[\rho_{k+1}] \otimes \ldots
  \otimes  \mathcal{T}^k[\rho_{k+m}]\quad \mbox{for} \quad m = 0, 1, 2, \ldots \ ; \\
{\rm (v)}&& \langle  \mathcal{T}^{(k)}[\rho_0 | \rho_1  \, |\, \ldots, \rho_k] \, | \,
         \widehat w^{\sigma}(\theta) \rangle_0
   =  \langle \rho_0 \, |\, \widehat w^{\sigma}(e^{ik\epsilon\tau}
   (g^{\sigma}z^{\sigma})^{k} \theta)  \rangle_0   \\
   && \times
       \prod_{j=1}^k  \langle \rho_j \, |\, \widehat w^{\sigma}(e^{ik\epsilon\tau}
   (g^{\sigma}z^{\sigma})^{k-j} g^{\sigma}w^{\sigma} \theta)  \rangle_0 .
\end{eqnarray*}
\end{lem}
{\sl Proof : }(i) This is obvious by definition.

\noindent(ii)  By Lemma \ref{RT}, we get
\[
     R_{0,k}T_k^{\sigma(k)}  T_{k-1}^{\sigma(k)}\ldots T_1^{\sigma(k)}
       = (R_{0,1}T_1^{\sigma(1)})( R_{1,2}T_1^{\sigma(2)}) \ldots (R_{k-1,k}T_1^{\sigma(k)}) .
\]
Then, definition (\ref{defTk}) and Lemma \ref{T[]}(vi) yield
\[
       \mathcal{T}^{(k)}[ \rho_0 | \rho_1, \ldots, \rho_k ]
       = (R_{0,1}T_1^{\sigma(1)})( R_{1,2}T_1^{\sigma(2)}) \ldots (R_{k-1,k}T_1^{\sigma(k)})
          (\rho_0 \otimes \rho_1 \otimes  \ldots \otimes \rho_k)
\]
\[
      = \mathcal{T}[  \mathcal{T}[ \ldots \mathcal{T}[  \mathcal{T}[ \rho_0 | \rho_1] |
       \mathcal{T} \rho_2] \ldots | \mathcal{T}^{k-2} \rho_{k-1}] | \mathcal{T}^{k-1} \rho_k],
\]
which iplies the claim.

\noindent(iii) This can be derived by induction using above (ii) and Lemma \ref{T[]}(iv).

\noindent(iv) Due to Lemma \ref{RT}, we have
\[
        R_{m,k+m}T_{k}^{\sigma(k+m)} T_{k-1}^{\sigma(k+m)} \ldots
           T_1^{\sigma(k+m)}(\rho_0 \otimes \rho_1 \otimes \ldots
     \otimes \rho_{k+m})
\]
\[
      =  \big(R_{m,m+1} T_1^{\sigma(m+1)}\big)
        \big(R_{m+1, m+2}T_1^{\sigma(m+2)} \big)
     \ldots
\]
\[
     \ldots \big(R_{m+k-1, m+k}T_1^{\sigma(m+k)}\big)
       (\rho_0 \otimes \rho_1 \otimes \ldots \otimes \rho_{k+m}) .
\]
Using successively Lemma \ref{T[]}(vi) and the result (ii) of the present lemma,
one obtains the assertion.

\noindent(v) By virtue of Lemma \ref{T[]}(ii) and of the result (i) above,
one can prove the case $k=1$.
Let us assume the validity for $k \geqslant 1$.
Then the validity of the case $k+1$ follows from the (ii) above and  the formula
\[
        \langle  \mathcal{T}^{(k+1)}[\rho_0 | \rho_1, \ldots, \rho_{k+1}]  \, |\,
         \widehat w^{\sigma}(\theta) \rangle_0
        = \langle  \mathcal{T}[\mathcal{T}^{(k)}[\rho_0 | \rho_1, \ldots, \rho_k]
          | \mathcal{T}^k\rho_{k+1}]  \, |\,
         \widehat w^{\sigma}(\theta) \rangle_0
\]

\vspace{-9mm}

\begin{eqnarray*}
     & = & \langle  \mathcal{T}^{(k)}[\rho_0 | \rho_1, \ldots, \rho_k]  \, |\,
         \widehat w^{\sigma}( e^{i\epsilon\tau}g^{\sigma}z^{\sigma} \theta) \rangle_0
     \, \langle  \mathcal{T}^k\rho_{k+1}  \, |\,
         \widehat w^{\sigma}( e^{i\epsilon\tau}g^{\sigma}w^{\sigma} \theta) \rangle_0
\\
    & = & \langle \rho_0 \, |\, \widehat w^{\sigma}(e^{ik\epsilon\tau} (g^{\sigma}
    z^{\sigma})^{k} e^{i\epsilon\tau} g^{\sigma}z^{\sigma} \theta)  \rangle_0
\end{eqnarray*}

\vspace{-5mm}

\[
      \times  \prod_{j=1}^k  \langle \rho_j \, |\, \widehat w^{\sigma}(e^{ik\epsilon\tau}
   (g^{\sigma}z^{\sigma})^{k-j} g^{\sigma}w^{\sigma}
         e^{i\epsilon\tau} g^{\sigma}z^{\sigma} \theta)  \rangle_0
       \langle \rho_{k+1} \, |\, \widehat w^{\sigma}(e^{i(k+1)\epsilon\tau}
    g^{\sigma}w^{\sigma} \theta)  \rangle_0 \,,
\]
which proves the assertion (v) by induction.   \hfill $\square$

Here, we comment that Lemma \ref{Tk[]} (v) is a revisit to the evolution of
the subsystem $\mathcal{S}$ in Lemma \ref{corr-m-step-evol}.
\subsection{Reduced density matrices of finite subsystems}
In this section, we consider evolution of subsystems $\mathcal{S}_{n, k}$ (\ref{S-nk}) and $\mathcal{S}_{\sim n}$.
Our aim is to study the large-time asymptotic behaviour of their states, when initial
density matrix is given by (\ref{state-S-C1}).

For the density matrix $\rho_1 $ in (\ref{state-S-C1}), we assume the condition:

\medskip

           [H] \qquad
     $ \displaystyle  D(\theta) = \prod_{l=0}^{\infty} \langle\rho_1 \, |\,
      \widehat w ( (g^{\sigma}z^{\sigma})^l\theta)\rangle_0 \, $
          converge for any          $ \, \theta \in \C$ \;

\noindent \hskip28mm
 and  the map $\, \R\ni t \mapsto D(t\theta) \in \C \;$
               is continuous.

\medskip

\noindent Here, we do not assume gauge invariance of $\rho_1$. (c.f. Theorem \ref{limit-S})

\noindent Under the condition [H], one obtains the following theorem:
\begin{thm}\label{limit-cavity}
There exists a unique density matrix $\rho_* $ on $ \mathscr{F} $ such that
$ \mathcal{T}[\rho_* \, |\,\rho_1] = \mathcal{T}\rho_* $ holds.
And $\rho_*$ also satisfies

\noindent {\rm (1)} \qquad
$\displaystyle
     \omega_{\rho_*}(\widehat w(\theta)) =
     \exp\Big[-\frac{|\theta|^2}{4}
    \frac{\sigma_- +\sigma_+}{\sigma_- - \sigma_+}
        \Big(1 -\frac{|g^{\sigma}(\tau)w^{\sigma}(\tau)|^{2}}
        {1 - |g^{\sigma}(\tau)z^{\sigma}(\tau)|^2 } \Big)\Big]
      D(g^{\sigma}(\tau)w^{\sigma}(\tau) \theta)
$;

\noindent {\rm (2)} \qquad
  $\displaystyle  \mathcal{T}^{(k)}[\, \rho_* |\,\rho_1,\ldots,
       \rho_1] = \mathcal{T}^k\rho_* $
  for $k>1$;

\noindent {\rm (3)} \qquad For any density matrix $ \rho_0 $ in (\ref{state-S-C1}),
the convergence
$ \; \lim_{k \to \infty}\displaystyle \mathcal{T}^{-k}\Big( \mathcal{T}^{(k)}[\rho_0 \, |
  \,  \rho_1, \ldots, \rho_1] \Big) = \rho_*  \; $
holds in the weak${}^*$-$\mathscr{A}^{(0)}$ topology on $\mathscr{C}^{(0)}$.
\end{thm}
\begin{remark}\label{weak*}
\noindent (a) The weak*-$\mathscr{A}^{(0)}$ topology on  $\mathscr{C}^{(0)}$
induced by the pair $(\mathscr{C}^{(0)}, \mathscr{A}^{(0)})$  (\ref{dualCA})
is coarser than the weak*-$\mathcal{L}(\mathscr{F})$ topology,  which coincides
with the weak and the norm topologies on the set of normal states  \cite{Ro, BR1}.

\noindent (b) When $\rho_1$ is gauge-invariant, the characteristic function
in (1) coincides with  (\ref{lim-m-expl}) and the present theorem
reduces to Theorem \ref{limit-S}.
Especially, the free evolution
$\mathcal{T}[\rho_* \, |\,\rho_1] = \mathcal{T}[\rho_*]$ reduces to the
invariance $\mathcal{T}[\rho_* \, |\,\rho_1] = \rho_*$.
\end{remark}
{\sl Proof }: First, we note that $\lim_{k \to \infty}\langle \rho_0|
w^{\sigma}( (g^{\sigma}z^{\sigma})^{k} \theta) \rangle_0 = 1$
because of $|g^{\sigma}z^{\sigma}| < 1$ and of the weak continuity of the state
$\omega_{\rho_0} = \langle \rho_0 | \, \cdot \, \rangle_0$.
Then by  Lemma \ref{Tk[]} (iii),(v) and Lemma \ref{T[]}(i), we get
\[
 \lim_{k\to\infty} \langle \mathcal{T}^{-k}\Big(\mathcal{T}^{(k)}[\rho_0 \, | \,
    \rho_1, \ldots, \rho_1]\Big) |\widehat w^{\sigma}(\theta)\rangle_0
\]
\begin{equation*}
 = \lim_{k\to\infty} \langle \mathcal{T}^{-k}[\rho_0] | \widehat w^{\sigma}(e^{ik\epsilon\tau}
   (g^{\sigma}z^{\sigma})^{k}  \theta) \rangle_0
\end{equation*}
\begin{equation} \label{stationary}
       \times      \prod_{j=0}^{k}  \langle \mathcal{T}^{-k}[\rho_1] | \widehat w^{\sigma} (e^{ik\epsilon\tau}
(g^{\sigma}z^{\sigma})^{k-j} g^{\sigma} w^{\sigma} \theta)  \rangle_0
\end{equation}
\[
          = \prod_{l=0}^{\infty}  \langle \rho_1 | \widehat w^{\sigma}((g^{\sigma}z^{\sigma})^{l} g^{\sigma}
    w^{\sigma} \theta)  \rangle_0
\]
\[
  = \exp\Big[\frac{|\theta|^2}{4}
    \frac{\sigma_- +\sigma_+}{\sigma_- - \sigma_+}
        \frac{|g^{\sigma}w^{\sigma}|^{2}}
        {1 - |g^{\sigma}z^{\sigma}|^2 } \Big]
            D(g^{\sigma} w^{\sigma} \theta) \ ,
\]
which means that $\displaystyle  \lim_{k\to\infty} \langle \mathcal{T}^{-k}\Big(\mathcal{T}^{(k)}[\rho_0 \, | \,
\rho_1, \ldots, \rho_1]\Big) | \widehat w^{\sigma}(\theta)\rangle_0  $ is equal
to the right-hand side of (1) in the theorem.  (Recall (\ref{m-W-m}) and (\ref{m-w}).)
The right-hand side of (1) satisfies:
(i) \textit{normalization}, (ii) \textit{unitarity} and (iii) \textit{positivity}, and (vi) \textit{regularity},
since it is a limit of characteristic functions, under condition [H].
Hence from the Araki-Segal theorem as in Section \ref{preliminaries},
there exists a state $\omega_*$ on the CCR-algebra $\mathscr{A}(\mathscr{F})$ such
that its characteristic function is given by the right-hand side of (1).
Moreover,
the continuity assumption about the function $D$ yields that  the
state $\omega_*$ is normal by the Stone-von Neumann uniqueness theorem \cite{BR2}.
Hence, there exists a density matrix $\rho_*$ such that
$\omega_* = \omega_{\rho_*}$, which conclude (1).
Now, (3) is obvious.

Free evolution $ \mathcal{T}[\rho_* \, |\,\rho_1] = \mathcal{T}\rho_* $ can be derived
from (1) by the use of
Lemma \ref{T[]} (iii),(i) and (\ref{m-W-m}), (\ref{m-w}). Indeed, one has
\[
  \langle \mathcal{T}[\rho_*|\rho_1]| \widehat w^{\sigma}(\theta) \rangle_0
= \langle \rho_* | \widehat w^{\sigma}
(e^{i\epsilon\tau}g^{\sigma}z^{\sigma}\theta) \rangle_0 \langle \rho_1 | \widehat w^{\sigma}
(e^{i\epsilon\tau}g^{\sigma}w^{\sigma}\theta) \rangle_0
\]
\[
 = \exp\Big[\frac{\sigma_-+\sigma_+}{4(\sigma_--\sigma_+)}(|g^{\sigma}
z^{\sigma}\theta|^2 + |g^{\sigma}w^{\sigma}\theta|^2)\Big]
\]
\begin{equation}\label{one-step-stationary}
 \times \langle \rho_* | \widehat w(e^{i\epsilon\tau}g^{\sigma}z^{\sigma}\theta) \rangle_0
\langle \rho_1 | \widehat w(e^{i\epsilon\tau}g^{\sigma}w^{\sigma}\theta) \rangle_0
\end{equation}
\[
 = \exp\Big[\frac{\sigma_-+\sigma_+}{4(\sigma_--\sigma_+)} \Big(|g^{\sigma}w^{\sigma}\theta|^2
+ \frac{|g^{\sigma}w^{\sigma}|^2} {1 - |g^{\sigma}z^{\sigma}|^2}|g^{\sigma}z^{\sigma}\theta|^2
\Big)\Big]
\]
\[
  \times D(g^{\sigma}w^{\sigma}
e^{i\epsilon\tau}g^{\sigma}z^{\sigma}\theta)
\langle \rho_1 | \widehat w(e^{i\epsilon\tau}g^{\sigma}w^{\sigma}\theta) \rangle_0
\]
\[
   = \exp\Big[\frac{\sigma_-+\sigma_+}{4(\sigma_--\sigma_+)}\frac{|g^{\sigma}w^{\sigma}|^2}
{1 - |g^{\sigma}z^{\sigma}|^2}
D(e^{i\epsilon\tau}g^{\sigma}w^{\sigma}\theta)
\]
\[
  = \exp\Big[\frac{\sigma_-+\sigma_+}{4(\sigma_--\sigma_+)} |\theta|^2\Big]
  \langle \rho_* | \widehat w
(e^{i\epsilon\tau}\theta) \rangle_0
= \langle \mathcal{T}[\rho_*] | \widehat w^{\sigma} (\theta) \rangle_0  =
\omega_{\mathcal{T}[\rho_*]}(\widehat w^{\sigma} (\theta))\ ,
\]
where we used the equality
$ D(g^{\sigma} z^{\sigma} \theta) \langle \rho_1 | \widehat w(\theta)  \rangle_0
 = D(\theta)$.

Now the assertion (2) follows directly from $\mathcal{T}[\rho_* \,|\,\rho_1] = \mathcal{T}\rho_* $, by
using Lemma \ref{Tk[]}(ii)(iii).

To prove the uniqueness of $\rho_*$, let $\rho_{\spadesuit}$ be another density matrix satisfying
$\mathcal{T}[\rho_{\spadesuit} \,|\,\rho_1] = \mathcal{T}\rho_{\spadesuit} $. Then, $\rho_{\spadesuit}$ satisfies the
property (2) and
\[
    \rho_{\spadesuit} = \lim_{k\rightarrow\infty} \mathcal{T}^{-k}[\mathcal{T}^{(k)}[\rho_0 \, |
  \,  \rho_1, \ldots, \rho_1]] \ ,
\]
which  coincides with $\rho_*$ by \textit{(3)}.
Hence, one gets $ \rho_{\spadesuit} = \rho_* $. \hfill $\square$

\medskip

Now we consider the large-time behaviour of the states
(\ref{S-rho-k}) of subsystems $\mathcal{S}_{\sim n}$.
Let $\rho_1$ be a density matrix on $\mathscr{F}$ satisfying the
condition [H]. Then we have the following theorem.
\begin{thm} \label{limit-subsystem}
For any density matrix $\rho_0$ on  $\mathscr{F}$ and $n, m \in \N, \; \red{m \geqslant n}$, the limit:
\[
   (\mathcal{T}^{-k})^{\otimes(m+1)} R_{m,m+k}T^{\sigma(m+k)}_{(n + k )\tau, 0}
    \big(\rho_0\otimes\rho_1^{\otimes (m+k)} \big)
       \longrightarrow T^{\sigma(m)}_{n\tau, 0}
    \big(\rho_*\otimes\rho_1^{\otimes m} \big)
   \;  \mbox{ as }  \; k\to\infty \ ,
\]
holds in the weak*-$\mathscr{A}^{(m)}$ topology on $\mathscr{C}^{(m)}$.
Here $\rho_* $ is the density matrix on $\mathscr{F}$ given
in Theorem \ref{limit-cavity}.
\end{thm}
{\sl Proof : } By Lemma \ref{RT}, Lemma \ref{Tk[]}(iv) and Lemma \ref{T[]}(iv), we obtain
\[
    (\mathcal{T}^{-k})^{\otimes(m+1)} R_{m,m+k}T^{\sigma(m+k)}_{(n + k )\tau, 0}
    \big(\rho_0\otimes\rho_1^{\otimes (m+k)} \big)
\]
\[
   = (\mathcal{T}^{-k})^{\otimes(m+1)} T^{\sigma(m)}_n \ldots T^{\sigma(m)}_1
R_{m,m+k}T^{\sigma(m+k)}_ k \ldots T^{\sigma(m+k)}_ 1
    \big(\rho_0\otimes\rho_1^{\otimes (m+k)} \big)
\]
\[
      = (\mathcal{T}^{-k})^{\otimes(m+1)} T^{\sigma(m)}_n \ldots T^{\sigma(m)}_1
       \big( \mathcal{T}^{(k)}[ \rho_0 | \rho_1, \ldots, \rho_1]\otimes
        (\mathcal{T}^k[\rho_1])^{\otimes m} \big)
\]
\[
      = T^{\sigma(m)}_n \ldots T^{\sigma(m)}_1 (\mathcal{T}^{-k})^{\otimes(m+1)}
      \big( \mathcal{T}^{(k)}[ \rho_0 | \rho_1, \ldots, \rho_1]\otimes (\mathcal{T}^k[\rho_1])^{\otimes m} \big)
\]
\[
      = T^{\sigma(m)}_{n\tau, 0}\big(\mathcal{T}^{-k}[\mathcal{T}^{(k)}
       [\rho_0 | \rho_1, \ldots, \rho_1]]\otimes \rho_1^{\otimes m}\big).
\]
Since one has
\[
\lim_{k \to \infty}\mathcal{T}^{-k}\Big(\mathcal{T}^{(k)} [\rho_0 | \rho_1, \ldots, \rho_1]\Big) = \rho_*
\]
in the weak*-$\mathscr{A}^{(0)}$ topology,
we obtain also the weak*-$\mathscr{A}^{(m)}$ convergence
\[
       (\mathcal{T}^{-k}[\mathcal{T}^{(k)}[\rho_0 \, | \, \rho_1, \ldots, \rho_1]])
       \otimes \rho_1^{\otimes m} \longrightarrow  \rho_*\otimes\rho_1^{\otimes m}
                  \quad \mbox{as} \  k \to \infty \, .
\]
By the duality (\ref{dualCA}), one also gets the continuity of $T_{n\tau,0}^{\sigma(m)}$ and hence, the weak*-$\mathscr{A}^{(m)}$
convergence
\[
       T^{\sigma(m)}_{n\tau, 0}\big(\mathcal{T}^{-k}[\mathcal{T}^{(k)}
       [\rho_0 \, | \, \rho_1, \ldots, \rho_1]]\otimes \rho_1^{\otimes m}\big)
           \longrightarrow  T^{\sigma(m)}_{n\tau, 0}
        \big( \rho_*\otimes\rho_1^{\otimes m} \big)
                  \quad \mbox{as} \ k \to \infty \, ,
\]
claimed in the theorem. \hfill$\square$

Let us put $m=n$ in the theorem. Then by (\ref{ptrace3}), we obtain the limit of the reduced density
matrix $\rho_{\mathcal{S}_{\sim n}}(\cdot)$ for the subsystem $\mathcal{S}_{\sim n}$:
\begin{cor} \label{limit-subsystem-corr}
The convergence
\begin{equation}\label{lim-subsys-corr}
 \lim_{k \to \infty} (\mathcal{T}^{-k})^{\otimes(n+1)}\rho_{\mathcal{S}_{\sim n}}((n+k)\tau)  =
T_{n\tau,0}^{\sigma(n)}(\rho_*\otimes\rho_1^{\otimes n})
\end{equation}
holds in the weak*-$\mathscr{A}^{(n)}$ topology on $\mathscr{C}^{(n)}$.
\end{cor}

Since $\mathcal{T}$ is the free evolution (\ref{T[-]}), the limit
(\ref{lim-subsys-corr}) means that dynamics of subsystem $\mathcal{S}_{\sim n}$ is the asymptotically-free
evolution of the state, which is given by the $n$-step evolution of the initial density matrix
$\rho_*\otimes\rho_1^{\otimes n}$ of the system $\mathcal{S} +\mathcal{C}_n$.

From the continuous time point of view, the subsystem $\mathcal{S}_{\sim n}$ shows the
asymptotic behaviour, which is a combination of the free evolution and the periodic evolution, c.f.
Remark \ref{Ex}(a).

\bigskip

\noindent
\textbf{Acknowledgements }

\noindent
H.T. thanks  JSPS for the financial support under the Grant-in-Aid for Scientific Research (C) 24540168.
He is also grateful to Aix-Marseille and Toulon Universities for their hospitality.
V.A.Z. acknowledges the Institute of Science and Engineering as well as Graduate School of the Natural Science and Technology
of Kanazawa University for hospitality and for financial support during the visit allowed to complete this paper.

\end{document}